# Gravitational Binding and Star Formation in Molecular Clouds of the Milky Way

Philip C. Myers[1], Mark Heyer[2], Ian W. Stephens[3,1], Simon Coudé[3,1], Nicole Karnath[4,1], and Howard A. Smith[1]

1. Center for Astrophysics | Harvard and Smithsonian, 60 Garden Street, Cambridge MA 02138 USA; pmyers@cfa.harvard.edu
2. Department of Astronomy, University of Massachusetts, Amherst, MA 01003, USA
3. Department of Earth, Environment, and Physics, Worcester State University, Worcester, MA 01602, USA
4. Space Science Institute, 4765 Walnut St., Ste. B., Boulder, CO 80301

**ABSTRACT**

The gravitational binding and star-forming properties of molecular clouds (MCs) in the Milky Way (MW) are estimated from CO cloud observations and from a model of pressure-bounded virial equilibrium (PVE). Two CO surveys are analyzed with the standard CO conversion factor. The main results are: (1) For each survey the cloud virial parameter $\alpha_{\mathrm{vir}}$ increases by a factor ~2 from galactocentric radius $R_{\mathrm{gal}} = 4$ kpc to 15 kpc. (2) PVE models match these trends only if the surface densities of survey clouds and nearby stars are comparable. This evidence of environmental influence resembles that seen in other disk galaxies. (3) Many survey clouds form stars even though their virial parameter exceeds the critical value $\alpha_{\mathrm{vir}} \approx 2$. In PVE such clouds with constant velocity dispersion have stable equilibrium and cannot form stars by simple global collapse. (4) However, simulations show that $\alpha_{\mathrm{vir}} \approx 2$ clouds with dissipating turbulence may form filaments, cores and protostars with little global contraction. Such clouds can match the MW star formation rate if their protostellar cores have a mass fraction $\sim 10^{-3}$. A simple model predicts that the star-forming age of a cloud is proportional to the ratio of its YSOs to its mass. (5) Clouds within 500 pc of the Sun are predicted to have star-forming ages 1-10 Myr and average YSO age ~2 Myr, matching evolutionary models. The Orion A cloud is predicted to have ~60 Class 0 protostars, ~2900 YSOs and efficiency $SFE \approx 0.02$, in good agreement with observed estimates.

## 1. Introduction

Our knowledge of star-forming molecular clouds began with the discovery of extended emission in the $J = 1 - 0$ rotational line of $^{12}$CO from the Orion Nebula (Wilson et al. 1970). Within a few years it became clear that CO emission from "Giant Molecular Clouds" (GMCs) is associated with H II regions and other markers of star formation throughout the Milky Way. There were more than ten large surveys of the Milky Way for CO emission between 1972 and 1980, and more than fifty between 1972 and 2005 (Heyer & Dame 2015; (HD15); see also Miville-Deschênes et al. 2017 (MD17)).

Once large samples of observed clouds were analyzed, power-law correlations were found between cloud velocity dispersion $\sigma_v$, radius $R$, and mass $M$ (Solomon et al. 1979, Larson 1981, Solomon et al. 1987). These "Larson relations" have been approximated as $\sigma_v \propto R^{1/2}$, $\sigma_v^2 \propto GM/R$, and mass surface density $\Sigma \approx$ constant from cloud to cloud. They are often ascribed to clouds with constant $\Sigma$ in "simple virial equilibrium" (SVE) where the internal kinetic energy is equal to half the gravitational energy (Field et al. 2011 (F11), Spitzer 1978, Bertoldi & McKee 1992 (BM92), McKee et al. 1993 (M93)).

The Larson relations may also be consistent with a more general virial system, where similar molecular clouds are surrounded by a medium of finite external pressure (Chièze 1987, Maloney 1988, Fleck 1988, Elmegreen 1989). The gravitational binding of clouds in such "pressure-bounded virial equilibrium" (PVE; F11) may range from strong binding with high pressure contrast, to weak binding with low pressure contrast. The external pressure on clouds is likely due to feedback from massive stars, and to the gravitational mass of gas and stars in the galactic environment.

In the midplane of a disk galaxy, star formation is expected to help set the external pressure on molecular clouds. In this picture, massive star feedback creates an extended layer of hot, diffuse gas. This hot gas occupies a large fraction of the volume of the interstellar medium (ISM); it is in approximate equilibrium with the warm and cool phases of the ISM. The midplane gas pressure $P_{\text{ext}}$ is then related to the surface densities of the disk gas $\Sigma_{\text{gas}}$ and of stars $\Sigma_{\text{star}}$ by $P_{\text{ext}} \approx (\pi/2)G\Sigma_{\text{gas}}(\Sigma_{\text{gas}} + \Sigma_{\text{star}})$ (Dopita 1985 eq. (2.6)). This equation approximates later derivations applied to CO observations of MCs, apart from factors of order unity (Elmegreen 1989, Blitz & Rosolowsky 2004, 2006, Leroy et al. 2008, Ostriker & Kim 2022 (OK22)).

These developments have led to a disk galaxy model of "dynamical equilibrium" (DE; OK22, Schinnerer & Leroy 2024). The average midplane pressure on a cloud at a given galactocentric radius is exerted by the local interstellar medium (ISM) gas. This pressure has a "gas gravity" component due to the self-gravitating layer of MW disk gas and an "external gravity" component due to nearby field stars. These "nearby" stars are in a thin midplane layer external to the cloud, whose vertical thickness equals that of the cloud. Their net gravitational pull on the ISM gas toward the midplane exerts a pressure on the cloud which may exceed that due to gas gravity alone. Henceforth we use "nearby" to describe the stars which contribute significantly to DE cloud pressure.

In the DE model the average rate of diffuse gas heating also equals the rate of diffuse gas cooling. The rates are regulated by massive star feedback and molecular cloud formation (Ostriker et al. 2010, Hopkins et al. 2011, OK22). The DE model is supported by observations of *PHANGS* disk galaxies (Sun et al. 2022) with resolution ~100 pc, and by MHD numerical simulations summarized in OK22. As described in OK22, DE is a more general version of PVE. In PVE, the cloud external pressure is a hydrostatic pressure on a cloud "surface." In DE, the cloud external pressure is a steady-state average. In this paper we use "DE" to refer to model fitting of the trends of virial parameter with galactocentric radius in Sections 2-4 and "PVE" to refer to the dynamical interpretation of virial parameter values in Section 5.

This paper compares the DE relations to cloud properties on finer scales than in extragalactic studies, by analyzing observations of clouds in the MW. These observations have typical cloud resolutions of 2-12 pc and sampling of the galactocentric radius $R_{\rm gal}$ at intervals of ~ 1 kpc. These comparisons test the level of agreement between the DE model and MW cloud observations despite statistical and observational uncertainties. They reveal how closely a cloud's gravitational binding is related to the incidence of its young stars. Similar estimates were made using earlier CO observations (e.g., Blitz & Rosolowsky 2004, 2006). However they did not cover the range of $R_{\rm gal}$ considered here, nor did they have a sufficient number of cloud observations to reach the statistical conclusions which are now possible.

In this work we use the HD15 and MD17 CO surveys of the MW, described in Section 3. We use the DE equations of vertical equilibrium, based on gas and star surface densities over a wide range of $R_{\rm gal}$. We express observed and model properties in terms of the MC virial parameter $\alpha_{\rm vir}$ (BM92) and in terms of the ratio of cloud internal and external pressure $P_{\rm int}/P_{\rm ext}$. The main result is that in each CO survey, MW clouds show a factor of ~2 increase of $\alpha_{\text{vir}}$ with $R_{\text{gal}}$ over $R_{\text{gal}} = 4 - 15$ kpc. This trend indicates that the self-gravitational binding of MCs becomes weaker from the inner to the outer galaxy. The increase of $\alpha_{\text{vir}}$ with $R_{\text{gal}}$ is consistent with balance between midplane pressure and gravitational weight, provided the weight is due primarily to nearby stars, whose surface density declines with $R_{\text{gal}}$.

In this paper, Section 2 describes the DE equations, assumptions, and method of analysis. Section 3 describes the CO observational data and how they are used. Section 4 describes fitting of the DE model to $\alpha_{\text{vir}}(R_{\text{gal}})$ for each CO survey, and implications of the fits for cloud properties and cloud binding. Section 5 discusses star-forming clouds which have high values of $\alpha_{\text{vir}}$. It presents a model which accounts for the galactic star formation rate by the collapse of dense cores within clouds which do not collapse globally. It shows agreement between model predictions and observations of star-forming clouds in the Solar neighborhood and in radial rings from 4 to 14 kpc. Section 6 discusses the limitations and implications of the results, and Section 7 presents a summary and conclusions.

## 2. Virial parameter of a pressurized molecular cloud

This section presents expressions for the virial parameter $\alpha_{\text{vir}}$ of a uniform spherical cloud and for the ratio $P_{\text{ext}}/P_{\text{int}}$ of its external pressure $P_{\text{ext}}$ and mean internal pressure $P_{\text{int}}$. The virial parameter of the cloud is defined as the ratio of twice the cloud internal kinetic energy to the magnitude of its self-gravitational energy, or

$$\alpha_{\text{vir}} \equiv \frac{5\sigma_v^2 R}{GM} \quad (1)$$

for a uniform cloud of mass $M$, radius $R$, and velocity dispersion $\sigma_v$ (BM92 eq. (2.8a)). The total external pressure $P_{\text{ext}}$ on midplane clouds due to a surrounding layer of gas with mass surface density $\Sigma_{\text{gas}}$ and a surrounding layer of field stars with mass surface density $\Sigma_{\text{star}}$ is approximated by

$$P_{\text{ext}} \approx \frac{\pi G \Sigma_{\text{gas}}^2}{2} + \Sigma_{\text{gas}} \left(\frac{G\Sigma_{\text{star}}}{z_{\text{star}}}\right)^{1/2} \sigma_{\text{eff}} \quad (2)$$

where $z_{star}$ is the scale height of the stellar disk, $\sigma_{eff} = (P_{ext}/\rho_0)^{1/2}$ is the effective velocity dispersion, and where $\rho_0$ is the midplane gas density (OK22 eq. (7)). Derivation of this equation is similar to that of eq. (11) of Elmegreen (1989) and to eq. (1) of Blitz & Rosolowsky (2004), with approximation uncertainty 10-20%. Eq. (2) neglects the contribution of dark matter to the midplane pressure. Its midplane density is ~6% of the gas density, when averaged over the range of galactocentric radii 4-16 kpc (OK22 Table 1). Therefore its contribution is negligible compared to other uncertainties in the DE model and in the observational data.

For a uniform cloud with constant velocity dispersion the internal pressure is $P_{int} = \rho \sigma_v^2$ where $\rho$ is the cloud mean density and $\sigma_v$ is the intensity-weighted velocity dispersion. $P_{int}$ can then be expressed in terms of $\alpha_{vir}$ and the mean cloud surface density $\Sigma_{cld}$ as

$$P_{int} = \frac{3\pi G}{20} \alpha_{vir} \Sigma_{cld}^2 \quad . \tag{3}$$

The virial parameter is also related to $P_{ext}/P_{int}$ for a pressurized uniform spherical cloud threaded by a uniform poloidal magnetic field. When the cloud mass is close to its maximum mass in virial equilibrium,

$$\alpha_{vir} \approx (1 - P_{ext}/P_{int})^{-1}(1 - \lambda^{-2}) \tag{4}$$

where $\lambda$ is the ratio of cloud mass to magnetic flux, normalized by its critical value between gravity and magnetic pressure (McKee et al. 1993 (M93), eqs. (9) and (19)). Equation (4) is applicable only for spheroidal clouds which are magnetically supercritical, since the requirement $\alpha_{vir} > 0$ implies $\lambda > 1$. For an infinitely long cylinder, virial equilibrium relations are given by Fiege & Pudritz (2000; FP00) and Li et al. (2022).

Equations (1) - (4) can be combined to express $\alpha_{vir}$ in terms of $\lambda$ and of ratios of surface densities,

$$\alpha_{vir} = 1 - \lambda^{-2} + \frac{10}{3}\left[\left(\frac{\Sigma_{gas}}{\Sigma_{cld}}\right)^2 + \left(\frac{\Sigma_{gas}}{\Sigma_{cld}}\right)\frac{(\Sigma_{star}\Sigma_0)^{1/2}}{\Sigma_{cld}}\right] \tag{5}$$

where $\Sigma_0 \equiv (2\sigma_{\text{eff}}/\pi)^2/(Gz_{\text{star}})$. This surface density $\Sigma_0$ is proportional to the surface density of a self-gravitating layer with velocity dispersion $\sigma_{\text{eff}}$ and scale height $z_{\text{star}}$. We adopt $z_{\text{star}} = 280$ pc based on recent *GAIA* estimates (Vieira et al. 2023). For $\sigma_{\text{eff}}$ we use the observed velocity dispersion of H I gas as in Elmegreen (1989), Blitz & Rosolowsky (2006), and Lada et al. (2025). We adopt the average value $\sigma_{\text{eff}} = 9$ km s$^{-1}$ based on H I observations in the LAB surveys (Kalberla et al. 2005; Marasco et al. 2017). Variation of $\sigma_{\text{eff}}$ with $R_{\text{gal}}$ has relatively small effect on model fit parameters, as discussed in Section 4.2.1. With these values of $z_{\text{star}}$ and $\sigma_{\text{eff}}$, $\Sigma_0 = 27\ M_\odot\ \text{pc}^{-2}$. The cloud pressure ratio $P_{\text{int}}/P_{\text{ext}}$ is then obtained from equations (4) and (5).

Values of $\alpha_{\text{vir}}$ from equation (1) based on CO observations (filled circles and error bars) will be compared with best-fit DE model curves of $\alpha_{\text{vir}}$ from equation (5) (smooth curves) in Section 4.

## 3. CO survey data

Numerous surveys of the MW have been made in the $J = 1 - 0$ rotational transition of $^{12}$CO, summarized in HD15 and compared in Evans et al (2021; E21). We have chosen two independent surveys for analysis, each with the greatest available sky coverage at its angular resolution. It gives a more complete picture to analyze both surveys, because they probe different spatial scales with different resolution, and because they use different methods of analysis.

Each "survey" is a combination of unbiased regional surveys, as detailed in Dame et al. (2001), HD15, and MD17. The "HD15" survey is a collection of studies made with the 14-m telescope of the Five College Radio Astronomy Observatory (FCRAO) with FWHM beam width 47 arcsec. They are summarized in HD15, Solomon et al. (1987), Heyer et al. (2001), and Roman-Duval et al. (2010). The "MD17" survey is based on observations made with 1.2-m telescopes at Columbia University, at the Harvard-Smithsonian Center for Astrophysics, and at the University of Chile with FWHM beam widths 8.4 and 8.6 arcmin (Dame et al. 2001). The HD15 survey has finer angular resolution by a factor ~11, while the MD17 survey covers a greater fraction of the MW plane. At a typical cloud distance of 5 kpc from the Sun, the HD15 survey beam width is 1.1 pc and the MD17 survey beam width is 12 pc. A typical MD17 cloud has mean density $\bar{n} \approx 25$ cm$^{-3}$ (MD17) while the most extensive HD15 survey has $\bar{n} \approx 100$ cm$^{-3}$ (Solomon et al. 1987).

In the analysis of HD surveys, a cloud is generally defined as a closed surface in two space coordinates and in spectral line velocity. Its plane-of-sky boundary encloses contiguous sky positions whose peak CO line brightness exceeds a minimum value, chosen to reduce cloud blending. In Solomon et al. (1987) this minimum brightness temperature is usually 4 K. More detailed methods of cloud segmentation are described in HD15.

In the MD17 analysis of the Dame et al. (2001) surveys, each CO spectrum is first fitted with multiple Gaussian functions having independent velocity, brightness, and velocity dispersion. Then clouds are identified as coherent structures using a hierarchical cluster analysis scheme similar to *clumpfind* (Williams et al. 1994). A value of integrated CO line intensity is chosen to define an initial candidate cloud boundary. As this threshold is reduced, emission outside the old boundary is evaluated, to be added to the old cloud, to join a neighbor cloud, or to define the start of a new cloud. This technique was able to segment 98% of the CO emission in Dame et al. (2001) into clouds. Further details are in MD17 Appendices A and B.

Each survey provides estimates of molecular cloud surface density $\Sigma_{cld}$ and virial parameter $\alpha_{vir}$. These estimates are based on the standard conversion factor from $^{12}$CO line integrated intensity to molecular gas surface density, $X_{CO} = 2.0 \times 10^{20}$ cm$^{-2}$ (K km s$^{-1}$)$^{-1}$ (Bolatto et al. 2013, HD15). The effect of variation in $X_{CO}$ with galactocentric radius $R_{gal}$ is discussed in Section 6.1.5. Each CO survey provides useful estimates of $\Sigma_{cld}$ and of $\alpha_{vir}$ as a function of galactocentric radius $R_{gal}$, from $R_{gal} = 4$ to 15 kpc in bins of $\Delta R_{gal} = 1$ kpc. The mean number of clouds in a 1 kpc bin is ~30 (HD15) and ~400 (MD17).

The MD17 and HD15 survey differences in angular resolution and in cloud definition lead to significantly different estimates of $\Sigma_{cld}$ and of the virial parameter $\alpha_{vir}$ at each $R_{gal}$. In Section 4 we compare their derived properties with each other and with DE model predictions.

## 4. Fitting dynamical equilibrium models to CO observations of Milky Way clouds

### 4.1. Surface densities

Applying the DE model in eq. (5) to observations requires the values of $\Sigma_{star}$, $\Sigma_0$, $\Sigma_{cld}$, and the ratio $\Sigma_{gas}/\Sigma_{cld}$. Among these quantities, $\Sigma_0$ is a fixed parameter defined in Section 2, and $\Sigma_{gas}/\Sigma_{cld}$ is a free parameter, assumed to follow $\Sigma_{gas}/\Sigma_{cld} < 1$, independent of $R_{gal}$. Its value is set by fitting the DE model to each observed trend of $\alpha_{vir}(R_{gal})$, for each CO survey. The remaining quantities in eq. (5) decrease with increasing $R_{gal}$. They are $\Sigma_{star}(R_{gal})$ from McMillan 2017 (McM17), $\Sigma_{cld}(R_{gal})$ from MD17, and $\Sigma_{cld}(R_{gal})$ from HD15, shown in Figure 1.

In Figure 1, the HD15 fit curve is $\Sigma_{cldHD15} = \Sigma_{cldHD15,0} \exp(-R_{gal}/R_{HD15,0})$ where $\Sigma_{cldHD15,0} = 560 \, M_\odot \, pc^{-2}$ and $R_{HD15,0} = 4.3$ kpc. The MD17 curve is $\Sigma_{cldMD17} = \Sigma_{cldMD17,0} \exp(-R_{gal}/R_{MD17,0})$ where $\Sigma_{cldMD17,0} = 195 \, M_\odot \, pc^{-2}$ and $R_{MD17,0} = 3.5$ kpc. The McM17 curve is a fit to the sum of the thick-disk and thin-disk expressions in McM17 Table 2, given by $\Sigma_{star} = \Sigma_{star,0} \exp(-R_{gal}/R_{star,0})$ where $\Sigma_{star,0} = 1040 \, M_\odot \, pc^{-2}$ and $R_{star,0} = 2.65$ kpc. The Pearson correlation coefficient of each fit is $> 0.9$. The range of $R_{gal}$ covers the MW disk. It excludes gas with $R_{gal} \lesssim 4$ kpc, where the HD15 data have too little coverage for a useful comparison.

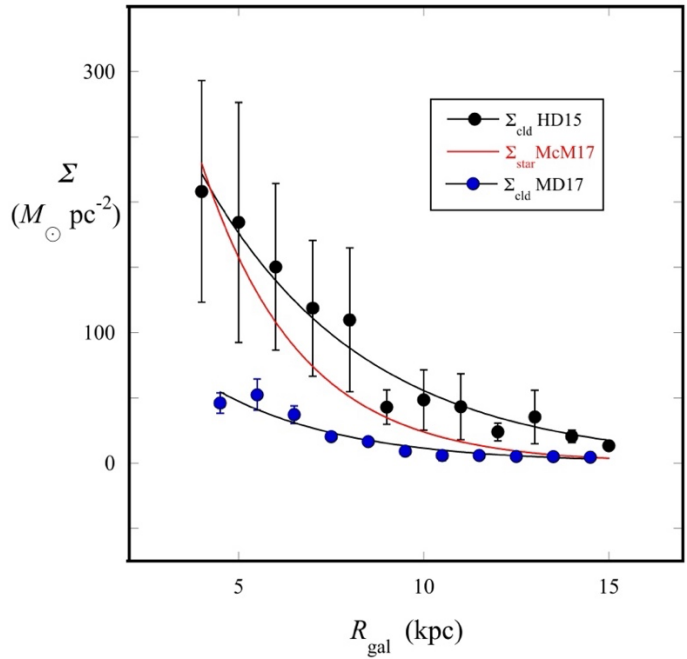

**Figure 1.** Mass surface densities of HD15 clouds (*black circles*), MD17 clouds (*blue circles*), and McM17 stars (*red curve*) as functions of galactocentric radius $R_{gal}$. The HD15 $\Sigma_{cld}$ data points

are mean values of surface density in 1 kpc bins. Each error bar represents the standard deviation about the mean. The MD17 $\Sigma_{cld}$ data points are medians in 1 kpc bins. MD17 do not present uncertainties in cloud surface density; error bars shown here are residuals after subtraction of the best-fit curve. Each best-fit curve is an exponential fit to the source data, to be used in the model fits in Section 4.2.

The fits in Figure 1 suggest that for each CO survey $\Sigma_{cld}$ declines with $R_{gal}$, as was noted by Heyer et al. (2009), HD15 and by MD17. This decline contrasts with the idea that $\Sigma_{cld}$ is constant, based on observations over a sample of local clouds (Larson 1981). Each decline in Figure 1 is well fit by an exponential profile. Exponential profiles of surface density with $R_{gal}$ are known for stars, molecular gas, and the star formation rate (Bigiel et al. 2008, Wang et al. 2019). In the MW, declining exponentials are used to describe the surface densities of stars and gas (McM17, MD17) over the range 4-14 kpc (Sysoliatina & Just 2022).

The physical origin of declining galactic profiles of molecular gas surface density with galactocentric radius $R_{gal}$ is not fully understood. In recent models the MW is described as an accretion disk whose viscosity arises from magnetic stresses (Wang and Lilly 2022). For MW surface densities derived from CO observations, some of the surface density decline may arise from assuming $X_{CO}$ to be constant while ignoring the decrease in the metallicity of the gas with $R_{gal}$ (Sodroski et al. 1995, Lada & Dame 2020, E21, and Evans et al. 2022; E22). However the decrease in metallicity may not account for the full range of surface density decline, as discussed in Section 6.1.5. The declining surface density of molecular cloud gas with $R_{gal}$ is a well known property of disk galaxies, reflecting influence of their galactic environment (S20, Schinnerer & Leroy 2024).

Figure 1 shows that the MD17 and HD15 surveys have significantly different mean values of cloud surface density $\Sigma_{cld}$, as noted by MD17 Section 4.1.2 and E21. The increase of surface density from MD17 clouds to HD15 clouds can be quantified by the ratio of exponential fit surface densities. The ratio of fit values $\Sigma_{cldHD15,fit}(R_{gal})/\Sigma_{cldMD17,fit}(R_{gal})$ has mean ± standard deviation 4.8 ± 0.9 when sampled over 0.5 kpc intervals of $R_{gal}$. This systematic increase from MD17 data to HD15 data may arise from a difference in cloud definition, since the line brightness threshold for a cloud map boundary is fainter for a MD17 cloud than for a HD15 cloud. Thus a MD17 cloud map boundary encloses faint emission from the periphery of the corresponding HD15 cloud map boundary. In addition, the 8.4 arcmin beam of the Dame et al. (2001) survey may smooth out small scale structure in the clouds that are more fully resolved with the 47 arcsec beam of the FCRAO surveys. This smoothing leads to lower observed antenna temperatures and lower mass surface densities. A comparison of clouds identified from the Dame et al. (2001) data by dendrogram analysis and by the MD17 method found that about 2500 of the 8100 MD17 clouds were not credibly detected (Lada & Dame 2020).

In Figure 1, the trend of $\Sigma_{\text{cld}}$ with $R_{\text{gal}}$ is better defined in the MD17 data than in the HD15 data. The MD17 survey has a statistical advantage since it samples an average of ~400 clouds in each ring of $R_{\text{gal}}$ rather than ~30 clouds in the HD15 survey.

### 4.2. Variation of virial parameter with galactocentric radius

#### 4.2.1. Trends and models of $\alpha_{\text{vir}}(R_{\text{gal}})$

The observed trend of $\alpha_{\text{vir}}(R_{\text{gal}})$ and its best-fit DE model fits based on equation (5) are shown in Figure 2 for MD17 data, and in Figure 3 for HD15 data. The fits are limited because reliable estimates of the mass-to-magnetic flux ratio in CO survey clouds are not available. However, Zeeman measurements of clouds having similar column densities indicate $\lambda \approx 2 - 3$ (Crutcher 2012). It is therefore assumed that the $\lambda^{-2}$ term can be neglected in fits of equation (5) to $\alpha_{\text{vir}}(R_{\text{gal}})$. Subcritical clouds with $\lambda < 1$ are discussed in Section 6.6.1.

For each set of survey data in Figures 2 and 3, $\alpha_{\text{vir}}(R_{\text{gal}})$ increases by a factor ~2 over the range $4 - 15$ kpc. In each case this rising trend of $\alpha_{\text{vir}}(R_{\text{gal}})$ is fit by the DE model, with correlation coefficient 0.76 (MD17) and 0.70 (HD15). The DE model matches the observed trend only when the local surface density of nearby stars $\Sigma_{\text{star}}(R_{\text{gal}})$ is significant. In contrast, the model fails to match the observations when the effect of nearby stars is removed.

These properties are shown in Figures 2 and 3 by the horizontal line "gas and no stars" where $\Sigma_{\text{star}}$ is set to zero in eq. (5), and by the horizontal line "no gas no stars" where $\Sigma_{\text{star}}$ and $\Sigma_{\text{gas}}$ are each set to zero in eq. (5). In each of these cases the DE model value of $\alpha_{\text{vir}}(R_{\text{gal}})$ is constant, and it does not fit the observed trend of $\alpha_{\text{vir}}(R_{\text{gal}})$. In the case "no gas no stars" the DE model cloud has no surrounding gas layer and no nearby stars, so it has zero external pressure. Thus $\alpha_{\rm vir}(R_{\rm gal}) = 1$ as expected for an isolated cloud in SVE.

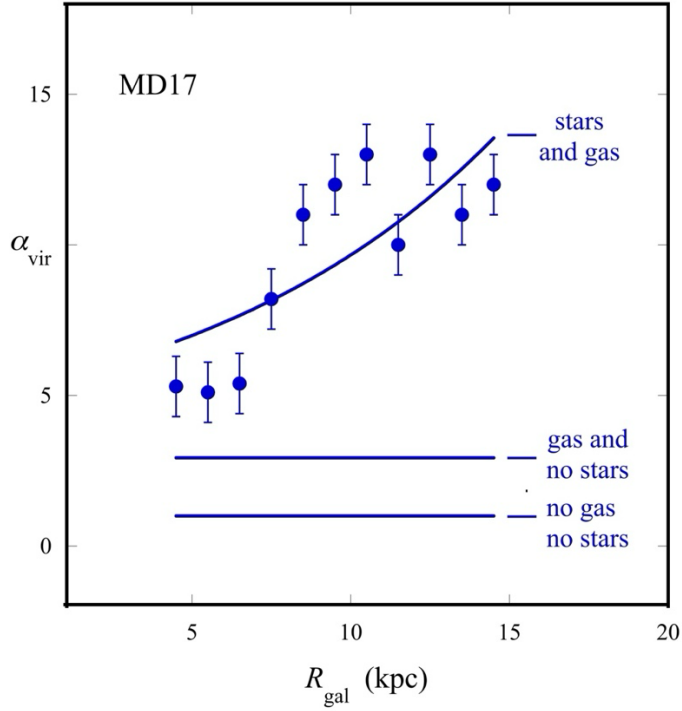

**Figure 2.** Median observed virial parameter $\alpha_{\rm vir}$ in equation (1) in rings of galactocentric radius $R_{\rm gal}$ and width $\Delta R_{\rm gal} = 1$ kpc, from CO observations analyzed by MD17 (*filled circles*). Error bars are upper limits on observed $\alpha_{\rm vir}$ estimated from MD17 Figure 16. The curve "stars and gas" is the best-fit DE model of $\alpha_{\rm vir}$ in eq. (5). The inputs $\Sigma_{\rm star}$ and $\Sigma_{\rm cldMD17,0}$ are taken from the exponential fits in Figure 1. The fixed parameter $\Sigma_0$ equals 27 $M_\odot$ pc$^{-2}$ based on $\sigma_{\rm eff} = 9$ km s$^{-1}$. The best-fit parameters are $\Sigma_{\rm gas}/\Sigma_{\rm cld} = 0.8 \pm 0.1$ and $R_{\rm MD17,0} = 3.4 \pm 0.3$ kpc with correlation coefficient 0.8. The line "gas and no stars" is the predicted DE value of $\alpha_{\rm vir}$ when $\Sigma_{\rm star} = 0$, and the line "no gas" is the predicted DE value when $\Sigma_{\rm star} = \Sigma_{\rm gas} = 0$.

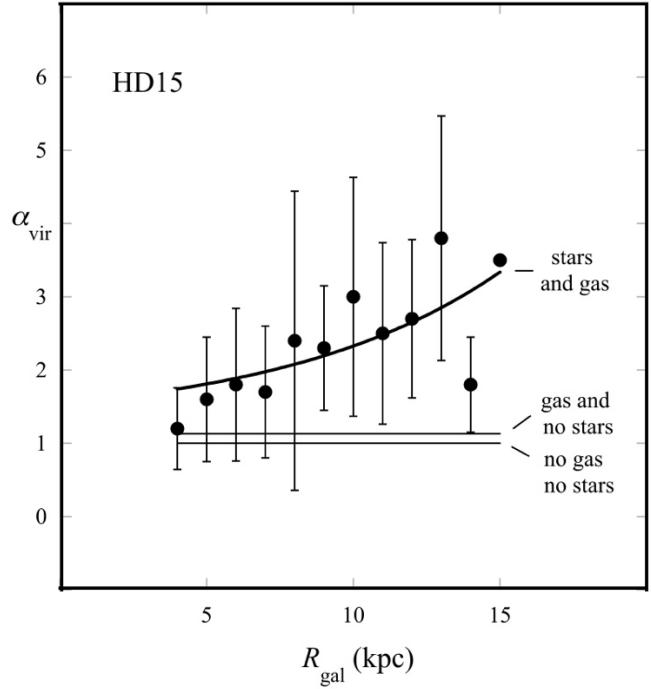

**Figure 3.** Mean observed virial parameter $\alpha_{\rm vir}$ from equation (1) in rings of galactocentric radius $R_{\rm gal}$ and width $\Delta R_{\rm gal} = 1$ kpc, from CO observations in HD15 (*filled circles*). Each error bar is the standard deviation of $\alpha_{\rm vir}$ values within the ring of radius $R_{\rm gal}$. The curve "stars and gas" is the best-fit DE model of $\alpha_{\rm vir}$ in eq. (5). The inputs $\Sigma_{\rm star}$ and $\Sigma_{\rm cldHD15,0}$ are from the exponential fits in Figure 1. The fixed DE model parameter $\Sigma_0$ equals 27 $M_\odot$ pc$^{-2}$ based on $\sigma_{\rm eff} = 9$ km s$^{-1}$. The best-fit DE model parameters are $\Sigma_{\rm gas}/\Sigma_{\rm cld} = 0.28 \pm 0.08$ and $R_{\rm HD15,0} = 3.1 \pm 0.3$ kpc, with correlation coefficient 0.7. The line "gas and no stars" is the predicted DE value of $\alpha_{\rm vir}$ when $\Sigma_{\rm star} = 0$, and the line "no gas" is the predicted DE value of $\alpha_{\rm vir}$ when $\Sigma_{\rm star} = \Sigma_{\rm gas} = 0$.

The fits of the DE model to the CO survey data in Figures 2 and 3 indicate that CO survey clouds are consistent with pressure-confined virial equilibrium. This conclusion is similar to results of a study of MW clouds in the "galactic ring" based on $^{13}$CO 1-0 observations (Keto 2025; Roman-Duval et al. 2009).

The importance of the stellar contribution to the DE model fits in Figures 2 and 3 can be understood from the relative contributions of the stellar and gas terms to the cloud external pressure $P_{\rm ext}$ in equation (2). For the adopted parameter values $\sigma_{\rm eff} = 9$ km s$^{-1}$ and $z_{\rm star} = 280$ pc, the stellar and gas contributions to $P_{ext}$ in equation (2) have the ratio $\left(P_{ext,stars}/P_{ext,gas}\right) =$ $6.5[\Sigma_{stars}/(M_\odot\ pc^{-2})]^{1/2}[\Sigma_{gas}/(M_\odot\ pc^{-2})]^{-1}$. At $R_{gal} = 10$ kpc, the DE fit to MD17 data indicates $P_{ext,stars}/P_{ext,gas} = 3.5$ while the fit to HD15 data indicates $P_{ext,stars}/P_{ext,gas} = 2.8$. For the simulation values of $\Sigma_{stars}$ and $\Sigma_{gas}$ in OK22 Table 1 over the range $R_{gal} \approx 2 - 16$ kpc, $P_{ext,stars}/P_{ext,gas}$ ranges from a minimum of 1.8 to a maximum of 3.8. Thus the stellar contribution to $P_{ext}$ exceeds the disk gas contribution by a factor of 2-4.

The fits of the DE model to observed values of $\alpha_{vir}$ for $R_{gal} = 4 - 15$ kpc assume that $\Sigma_{gas}$ and $\Sigma_{cld}$ decline similarly with $R_{gal}$, so that their ratio $\Sigma_{gas}/\Sigma_{cld}$ is approximately constant. This assumption was checked for the MD17 and HD15 data over the range of $R_{gal}$. We compared the value and uncertainty of the fit parameter $\Sigma_{gas}/\Sigma_{cld}$ to the mean and standard deviation of independent estimates of $\Sigma_{gas}/\Sigma_{cld}$. Here the independent estimate of $\Sigma_{gas}/\Sigma_{cld}$ is from the disk galaxy simulations of $\Sigma_{gas}$ in OK22, Table 2, and from the exponential fits to $\Sigma_{cld}$ in Figure 1. For the MD17 data the DE fit estimate of $\Sigma_{gas}/\Sigma_{cld}$ is $0.8 \pm 0.1$, while the independent values of $\Sigma_{gas}/\Sigma_{cld}$ have mean ± standard deviation $0.7 \pm 0.4$. For the HD15 data the DE fit estimate is $\Sigma_{gas}/\Sigma_{cld} = 0.28 \pm 0.08$, while the independent values have $\Sigma_{gas}/\Sigma_{cld} = 0.15 \pm 0.02$. Although the statistics are limited, for each CO survey the assumption of constant $\Sigma_{gas}/\Sigma_{cld}$ appears consistent with independent estimates of $\Sigma_{gas}/\Sigma_{cld}$ over the range of $R_{gal}$ used for fitting.

The DE model fits in Figures 2 and 3 assume constant effective velocity dispersion $\sigma_{eff} = 9$ km s$^{-1}$ based on the LAB H I surveys (Kalberla et al. 2005) analyzed by Marasco et al. (2017). This analysis indicates that $\sigma_{eff}$ decreases from ~11 km s$^{-1}$ to ~8 km s$^{-1}$ over $R_{gal} = 3.0 - 8.3$ kpc, with mean ± standard deviation $\sigma_{eff} = 8.9 \pm 1.1$ km s$^{-1}$. We estimated the effect of changing $\sigma_{eff}$ on the DE model fit parameters $\Sigma_{gas}/\Sigma_{cld}$, $R_{MD17,0}$ and $R_{HD15,0}$ by performing DE model fits with $\sigma_{eff} = 10, 8,$ and 6 km s$^{-1}$. For each CO survey, decreasing $\sigma_{eff}$ from 10 to 8 km s$^{-1}$ and from 8 to 6 km s$^{-1}$ increased the fit value of $\Sigma_{gas}/\Sigma_{cld}$ by ~10% and decreased $R_{MD17,0}$ and $R_{HD15,0}$ by ~2%. This change in each fit parameter was less than the fit uncertainty in that parameter. Also the best-fit curve of $\alpha_{vir}$ had negligible change as $\sigma_{eff}$ was changed, and the correlation coefficient changed by $\leq 0.01$. Thus assuming $\sigma_{eff} \approx 9$ km s$^{-1}$ is an acceptable approximation for our DE model fitting.

### 4.2.2. Similarity of virial parameter trends

In Figure 2, the values of $\alpha_{\text{vir}}(R_{\text{gal}})$ from MD17 are systematically greater than those from HD15 in Figure 3, by a nearly constant factor of ~4. This factor can be attributed largely to the typical ratio of surface densities $\Sigma_{\text{cld}}(R_{\text{gal}})$ between the two surveys shown in Figure 1. The virial parameter $\alpha_{\text{vir}}$ depends inversely on $\Sigma_{\text{cld}}$ according to $\alpha_{\text{vir}} = 5\sigma_v^2/(\pi GR\Sigma_{\text{cld}})$ from eq. (1), assuming that the cloud mass is $M = \pi R^2 \Sigma_{\text{cld}}$. Thus if $\sigma_v^2/R$ varies with $R_{\text{gal}}$ by a smaller factor than does $\Sigma_{\text{cld}}(R_{\text{gal}})$ from the HD15 data to the MD17 data, one may expect a systematic increase by a factor ~4 in $\alpha_{\text{vir}}(R_{\text{gal}})$ from the HD15 data to the MD17 data.

More directly, this property can be ascribed to the relative beam sizes and the beam filling factor of the CO emission. When the filling factor is <1, the measured antenna temperature is less than if the filling factor is 1. Then the integrated intensity is reduced, so the surface density is reduced. The expected reduction in surface density is verified by the direct comparison of $\alpha_{\text{vir}}(R_{\text{gal}})$ profiles in Figure 4.

Figure 4 shows that the MD17 and HD15 trends of $\alpha_{\text{vir}}(R_{\text{gal}})$ have similar shape and a nearly constant ratio. The upper part of Figure 4 shows the best-fit MD17 curve and each MD17 data point, colored blue and labelled MD17 as in Figure 2. The lower part shows the MD17 data multiplied by the ratio of surface density fits 0.23, colored red and labelled MD17'. It shows the HD15 data multiplied by 1, colored black and labelled HD15 as in Figure 3. Then the red and black fit curves closely coincide, and the red and black data points approximate the same trend. A fit to the combined HD15 and MD17' data has a greater correlation coefficient than the fit to the HD15 data alone. This evidence of similarity suggests that the two independently observed trends reflect the same basic property of $\alpha_{\text{vir}}(R_{\text{gal}})$ in MW molecular clouds.

The shape of the $\alpha_{\text{vir}}(R_{\text{gal}})$ trend common to both surveys can be approximated by assuming that $\alpha_{\text{vir},0}(R_{\text{gal}})$ equals the best-fit model to the HD15 data in Figure 3. Then one can write

$$\alpha_{\text{virHD15}}(R_{\text{gal}}) \approx \beta_{\text{HD15}} \alpha_{\text{vir},0}(R_{\text{gal}}) \quad (6)$$

and

$$\alpha_{\text{virMD17}}(R_{\text{gal}}) \approx \beta_{\text{MD17}} \alpha_{\text{vir},0}(R_{\text{gal}}), \quad (7)$$

where $\beta_{\text{HD15}} = 1$, $\beta_{\text{MD17}} = 4.3$, and where the best-fit model is

$$\alpha_{\text{vir},0}(R_{\text{gal}}) = 1 + \left(\frac{10}{3}\right)\{a + b\exp[R_{\text{gal}}(c^{-1} - d^{-1})]\}, \qquad (8)$$

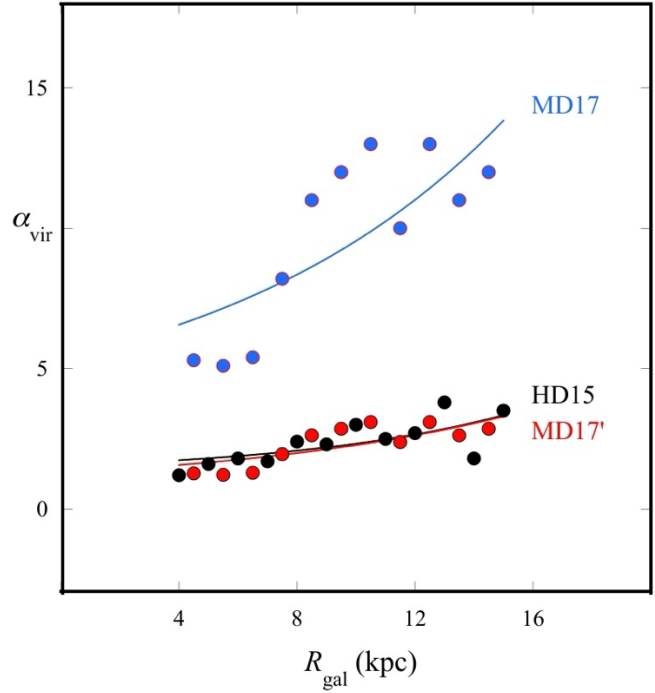

**Figure 4.** Trends of $\alpha_{\text{vir}}(R_{\text{gal}})$ for the MD17 and HD15 CO surveys. *Blue circles* show the observed $\alpha_{\text{vir}}$ data (equation 1) from MD17 in Figure 2, and the *blue curve* shows the best-fit DE model (equation 5) to the MD17 data in Figure 2. *Black circles* show the observed **$\alpha_{\text{vir}}$** data (equation 1) from HD15 in Figure 3, and the *black curve* shows the best-fit DE model (equation 5) to the HD15 data in Figure 3. For simplicity error bars are not shown. The *red curve* shows the best-fit DE model curve for MD17 data, multiplied by the mean fit ratio $r_\alpha = 0.23$, and the *red circles* show the MD17 data points multiplied by $r_\alpha$. Red symbols are labelled MD17' to indicate that they originate from MD17 data.

with $a = 0.08 \pm 0.05$, $b = 0.08 \pm 0.02$, $c = 3.1 \pm 0.3$ kpc, and $d = 5.3$ kpc. This relation indicates that inner galaxy clouds are more strongly self-gravitating than outer galaxy clouds, i.e. $\alpha_{\text{vir},0}(4 \text{ kpc}) = 1.7$ while $\alpha_{\text{vir},0}(15 \text{ kpc}) = 3.6$. The corresponding cloud pressure ratio decreases from inner galaxy to outer galaxy, with $P_{\text{int}}/P_{\text{ext}}(4 \text{ kpc}) = 2.3$ while $P_{\text{int}}/P_{\text{ext}}(15 \text{ kpc}) = 1.4$, based on equation (4).

The approximation $\alpha_{\text{vir},0}(R_{\text{gal}})$ in equation (8) is based on the DE fit to the HD15 data in Figure 3. In principle it could instead be based on the DE fit to the MD17 data in Figure 2, where each value of $\alpha_{\text{vir},0}(R_{\text{gal}})$ in equation (8) would then be increased by a factor $\approx 4$. However the HD15 fit appears preferable to the MD17 fit because its typical value of $\alpha_{\text{vir}}$ more closely approximates the typical value of $\alpha_{\text{vir}}$ in the *PHANGS-ALMA* galaxy survey, as described in Section 4.2.5.

### 4.2.3. Origin of virial parameter trends

In Figures 2, 3, and 4 the rising trends of the best-fit DE models of $\alpha_{\text{vir}}(R_{\text{gal}})$ are expected from the DE model eq. (5) and from the properties of its surface densities $\Sigma_{\text{star}}$ and $\Sigma_{\text{cld}}$. The rising trends of $\alpha_{\text{vir}}(R_{\text{gal}})$ are consistent with the comparable declines of $\Sigma_{\text{star}}$ and $\Sigma_{\text{cld}}$ with $R_{\text{gal}}$, shown in Figure 1.

A more specific condition for the increase of $\alpha_{\text{vir}}$ with $R_{\text{gal}}$ can be derived by assuming that $\Sigma_{\text{star}}$ and $\Sigma_{\text{cld}}$ are each declining exponentials with similar amplitudes and with respective scale lengths $R_{\text{star},0}$ and $R_{\text{cld},0}$. Then the right-hand term inside the square bracket of eq. (5) can be written $fg(R_{\text{gal}})$, where $f = (\Sigma_{\text{gas}}/\Sigma_{\text{cld}})(\Sigma_{\text{star},0}\Sigma_0)^{1/2}\Sigma_{\text{cld},0}^{-1}$ is a constant and where $g(R_{\text{gal}})$ has a simple dependence on $R_{\text{gal}}$,

$$g(R_{\text{gal}}) = \exp\left[\frac{R_{\text{gal}}}{R_{\text{cld},0}}\left(1 - \frac{R_{\text{cld},0}}{2R_{\text{star},0}}\right)\right] \quad . \tag{9}$$

Equation (9) is a rising exponential function of $R_{\text{gal}}$ provided $R_{\text{star},0} > R_{\text{cld},0}/2$, otherwise it is constant or declining. The fit curves in Figure 1 meet the rising exponential condition for the adopted values of $R_{\text{star},0}$ and for $R_{\text{cld},0}$.

Thus the increasing trend of $\alpha_{\rm vir}(R_{\rm gal})$ in MW clouds can be understood from (1) the approximate equilibrium between midplane gas pressure and the gravitational weight of nearby stars and gas, and from (2) the approximately exponential decline of star and cloud surface densities with $R_{\rm gal}$, provided that (3) the star surface density scale length is at least half the cloud surface density scale length.

**4.2.4. Line-width-size coefficient**

The values of $\alpha_{\rm vir}\Sigma_{\rm cld}$ in Section 4.2.1 can be used with equation (1) to estimate a key property of the Larson relations, the line-width-size scaling coefficient

$$(\sigma_v^2/R)^{1/2} = [(\pi G/5)\alpha_{\rm vir}\Sigma_{\rm cld}]^{1/2} \qquad (10)$$

(Heyer et al. 2009) as a function of $R_{\rm gal}$ for each CO survey. For each survey, $(\sigma_v^2/R)^{1/2}$ declines monotonically with similar shape from $R_{\rm gal} = 4$ to 15 kpc. When sampled over this range of $R_{\rm gal}$, $(\sigma_v^2/R)^{1/2}$ for the HD15 survey and for the MD17 survey have nearly equal means, 0.6 km s$^{-1}$pc$^{-1/2}$, consistent with MD17 Figure 13, and nearly equal standard deviations, 0.2 km s$^{-1}$pc$^{-1/2}$.

Although the observed surface density and virial parameter differ systematically from the HD15 survey to the MD17 survey, the line-width-size scaling coefficients derived from the two surveys are nearly equal, with difference in value less than 0.05 km s$^{-1}$pc$^{-1/2}$. This relatively small variation in $(\sigma_v^2/R)^{1/2}$ from the HD15 survey to the MD17 survey is consistent with the systematic increase in $\alpha_{\rm vir}(R_{\rm gal})$ from the HD15 survey to the MD17 survey, discussed in Section 4.2.2.

**4.2.5. Comparison with extragalactic studies**

A detailed analysis of CO observations in the *PHANGS-ALMA* disk galaxy survey indicates consistency with the DE model in OK22 and in associated papers (Sun et al. 2020 (S20); Sun et al. 2022 (S22)). The *PHANGS-ALMA* survey (Leroy et al. 2021) provided sensitive, high-resolution, wide field-of-view CO (2-1) imaging data for ~90 high-mass, star-forming galaxies.

The survey enabled systematic, quantitative studies of the relations between the properties of giant molecular clouds (GMCs) and their galactic environments. The galaxies were selected to be close enough, and the observational resolution was fine enough, to study these relations on cloud scales ≲ 150 pc. The CO data were analyzed "object-by-object" using a cloud segmentation algorithm, and also "pixel-by-pixel" where the beam size corresponds to the size of a typical GMC or giant molecular association.

The survey data show a well defined correlation between cloud internal turbulent pressure and midplane pressure due to galactic weight, with pressure ranging over $10^{4-7} k$ K cm$^{-3}$ where $k$ is Boltzmann's constant (S20). At the 120 pc size scale the typical pressure ratio is $P_{\text{int}}/P_{\text{ext}} = 2.8$. According to equation (4) this ratio corresponds to a virial parameter $\alpha_{\text{vir}} = 1.6$, a typical value for the inner MW in the lower part of Figure 4.

The values of $\alpha_{\text{vir}}$ in *PHANGS-ALMA* survey galaxies are more consistent with $\alpha_{\text{vir}}$ based on HD15 survey data than with $\alpha_{\text{vir}}$ based on MD17 survey data. An initial sample of 28 *PHANGS-ALMA* galaxies have a median $\alpha_{\text{vir}} = 2.7$ (S20), and a larger sample of 90 galaxies have $\alpha_{\text{vir}} = 1.1$ for object sampling or $\alpha_{\text{vir}} = 1.6$ for pixel sampling (S22). The HD15 surveys have median $\alpha_{\text{vir}} = 1.8$ (Solomon et al. 1987, Heyer et al. 2001), which lies within the range of *PHANGS-ALMA* values. In contrast the MD17 surveys have median $\alpha_{\text{vir}} = 8.5$ (MD17 Table 2), far outside the range of *PHANGS-ALMA* values.

This comparison of $\alpha_{\text{vir}}$ between the *PHANGS-ALMA* CO 2-1 observations and the HD15 CO 1-0 observations can be questioned because CO 2-1 observations are generally sensitive to denser gas than CO 1-0 observations (E21 Figure 12). However estimates of the median cloud density in the *PHANGS-ALMA* observations and in the HD15 surveys indicate density ≈ 100 cm$^{-3}$ in each case. These estimates are based on cloud properties in S22 Table 2 and in Solomon et al. (1987) Table 1. Their median densities may be similar because small dense regions are brighter in CO 2-1 than in CO 1-0 due to excitation, but they are fainter in *PHANGS-ALMA* observations than in HD15 observations due to beam dilution.

An analysis of 34 *PHANGS* galaxies evaluated the ratio of the gas and stellar terms in the DE pressure equation, as was done for MW clouds in Section 4.2.1 above. As was found for the MW clouds, the stellar pressure term dominates the gas pressure term by a typical factor ~2, with a slightly greater factor for weak spirals than for strong spirals (Elmegreen 2024, Figure 11).

In the Andromeda Galaxy (M31), CO 2-1 observations of molecular clouds with the Submillimeter Array reveal properties similar to those identified here for MW clouds. With synthesized beam size ~17 × 14 pc, a sample of 117 clouds has median $\alpha_{\rm vir}$ close to 2 (Lada et al. 2024 Figure 10; L24). This estimate is similar to typical values of $\alpha_{\rm vir}$ found in HD15 clouds and in the *PHANGS-ALMA* survey. This similarity supports the choice of $\alpha_{\rm vir,0}(R_{\rm gal})$ in equation (8) to resemble the HD15 trend in Figure 3 rather than the MD17 trend in Figure 2.

Also in M31, analysis with SVE and PVE models indicates that many observed clouds would be unbound according to SVE, whereas they appear pressure confined when their galactic environment is taken into account (Lada et al. 2025; L25; see also Keto et al. 2025). L25 also find that the confining pressure is due mainly to nearby stars. The stellar contribution to the total confining pressure is greater than that due to disk gas by a factor 2.1. This factor lies within the range of contribution ratios 2-4 found in Section 4.2.1 for MW clouds.

**4.2.6. Effect of cloud distance on estimates of virial parameter**

Since the virial parameter $\alpha_{\rm vir}$ varies inversely with cloud radius, one may ask how poor resolution of distant clouds is related to the trend of $\alpha_{\rm vir}$ with $R_{\rm gal}$ described above in Section 4. We find that differences in resolution and cloud boundary definition are needed to explain the systematic factor of ~4 ratio between HD15 and MD17 estimates of $\Sigma_{\rm cld}$ in Figure 1, and of $\alpha_{\rm vir}$ in Figure 4, as discussed in Section 4.1 and 4.2.2.

These resolution and definition differences appear to account for the factor of ~4 offset between the mean values of the HD15 and MD17 trends of $\alpha_{\rm vir}$ with $R_{\rm gal}$ in Figures 2, 3, and 4. But this factor of ~4 does not account for each trend shape, i.e. for the factor of ~2 increase in $\alpha_{\rm vir}$ from $R_{\rm gal} = 4$ kpc to 14 kpc, observed independently in the HD15 data and in the MD17 data. These trend shapes cannot be due to a simple dependence on cloud distance, since for each CO survey the cloud distances from the Sun and from the Galactic Center are poorly correlated. They cannot be due to simple beam dilution, since the survey beam areas differ by a factor of ~115. Instead we believe the trends of $\alpha_{\rm vir}$ with $R_{\rm gal}$ are likely consistent with gravitational pressure balance in the MW midplane, and with declining surface density of MW stars and gas with $R_{\rm gal}$, as discussed in Section 4.2.3.

# 5. Gravitational binding and star formation

The virial parameter $\alpha_{\rm vir}$ is often used to quantify the ability of a cloud to form stars. This section relates the observed and modeled values of $\alpha_{\rm vir}$ in Section 4 to observed and predicted rates of star formation. The values of $\alpha_{\rm vir}$ exceeding the critical value $\alpha_{\rm vir} \geq 2$ are interpreted to be inconsistent with simple global collapse of their mean-density gas. Section 5.3 describes an alternate way that such MCs can match the MW star formation rate, if they form stars by the local collapses of their dense cores.

## 5.1. Dynamical interpretation of virial parameter values

This section summarizes the interpretation of values of $\alpha_{\rm vir}$ for clouds in SVE and in PVE, for application to the properties of $\alpha_{\rm vir}(R_{\rm gal})$ described in Section 4.1. It also discusses simulations of clouds not in SVE or PVE, including gravoturbulent clouds, clouds in global hierarchical collapse, and clouds whose turbulence dissipates.

### 5.1.1. Simple virial equilibrium (SVE)

If an idealized cloud in virial equilibrium has negligible external pressure and magnetic field strength, its equilibrium is called simple virial equilibrium (SVE), where the internal kinetic energy equals half the gravitational energy (Spitzer 1978, BM92, F11). In SVE the cloud is "gravitationally bound" or "bound" or "virialized." The value of $\alpha_{\rm vir}$ in SVE is $\alpha_{\rm vir,S} = 1$ from eq. (1). Simple departures from uniform spherical structure, including increases in cloud central concentration, elongation, or velocity dispersion with radius, can increase $\alpha_{\rm vir,S}$ by a factor $\lesssim 2$ (BM92, MD17).

If $\alpha_{\rm vir} < \alpha_{\rm vir,S}$ the cloud has no possible equilibrium, so it undergoes "global collapse." If it is initially uniform, all of its mass falls radially inward from rest in a free-fall time based on its initial mean density (Hunter 1962). With small fluctuations, such global collapse is expected to promote fragmentation and star formation, since increasing density can decrease the gas temperature and decrease the Jeans mass (Hoyle 1953, Spitzer 1978). If instead $\alpha_{\rm vir} > \alpha_{\rm vir,S}$ the cloud is "unbound" since it has no external binding agent and its self-gravity cannot bind its internal motions. It is expected to expand on a cloud-crossing time scale at a speed equal to its velocity dispersion, preventing or inhibiting star formation.

Some authors adopt the condition $\alpha_{vir,S} = 2$ based on equating the internal kinetic and gravitational energy, in which case they are "bound" if $\alpha_{vir} < 2$ and "unbound" if $\alpha_{vir} > 2$ (E21).

### 5.1.2. Pressure-bounded virial equilibrium (PVE)

The critical stability of a cloud in PVE can be expressed in two ways. For a cloud with fixed mass and velocity dispersion, solutions to the PVE equation are unstable if the cloud radius is less than a critical length that is similar to the Jeans length; otherwise they are stable. Alternately, for fixed external pressure and velocity dispersion, PVE solutions are unstable if the cloud mass is greater than a critical mass similar to the Jeans mass; otherwise they are stable (M93, F11).

In the PVE of a uniform spherical cloud, the critical value of $\alpha_{vir}$ for gravitational binding is $\alpha_{vir,P} = 4/3$ based on equation (1) and on F11 equation (2). When $\alpha_{vir} = \alpha_{vir,P}$, the cloud external pressure has the maximum value which allows PVE, $P_{ext,max} = (\pi G/20)\Sigma_{cld}^2$. The value $\alpha_{vir,P} = 4/3$ is of the same order as $\alpha_{vir,S} = 1$ for SVE given above. However, the physical properties of the cloud gas differ between PVE and SVE for a given value of $\alpha_{vir}$.

Unstable PVE clouds with $\alpha_{vir} < \alpha_{vir,P}$ are interpreted as allowing collapse, fragmentation and star formation (F11). This outcome is similar to that of SVE clouds with $\alpha_{vir} < \alpha_{vir,S}$. However, clouds in PVE with $\alpha_{vir} > \alpha_{vir,P}$ are in stable equilibrium; they do not expand and disperse. F11 find that the CO survey observations of Heyer et al. (2009) are more consistent with clouds in PVE than in SVE. They require a range of cloud external pressures from $\sim 3 \times 10^4\ k\,{\rm K\,cm^{-3}}$ to $\sim 3 \times 10^6\ k\,{\rm K\,cm^{-3}}$ rather than a single value. They also find that stable PVE clouds in the Heyer et al. (2009) data are generally close to the critical point between the stable and unstable parts of the equilibrium curve (F11 Figure 3).

Since the critical values of $\alpha_{vir}$ lie in the range $\alpha_{vir} \approx 1 - 2$ for both $\alpha_{vir,S}$ and for $\alpha_{vir,P}$ we approximate them for simplicity as $\alpha_{vir,S} \approx \alpha_{vir,P} \approx 2$.

### 5.1.3. Star-forming turbulent PVE clouds

PVE clouds may be "gravoturbulent" (GT), where supersonic magnetized turbulence stabilizes the large-scale cloud gas against collapse. Its small-scale shocks compress gas which becomes gravitationally unstable against collapse and star formation (Krumholz & McKee 2005, Federrath & Klessen 2012). In "global hierarchical collapse (GHC)," self-gravity drives multiscale collapse similar to Hoyle fragmentation (Hoyle 1953, Ballesteros-Paredes et al. 2011, Vázquez-Semadeni et al. 2019). GT and GHC processes may be difficult to distinguish from observations because their characteristic velocities differ only by a factor $\sim\sqrt{2}$ when the virial parameter is $\alpha_{\text{vir}} \approx 2$ (Ballesteros-Paredes et al. 2011). They each can be described as "energy equipartition" models (Chevance et al. 2020).

A closely related process may occur when the initial supersonic turbulent motions of a nearly critical PVE cloud are allowed to dissipate on a crossing time scale. Simulations indicate that such a cloud has little initial global contraction, but it undergoes disordered turbulent collapse on smaller scales (Grudić et al. 2019 (G19), Grudić et al. 2021 (G21), Grudić et al. 2022 (G22), Offner et al. 2025 (O25)). The cloud develops multiple centers of infall, where protostellar dense cores form in contracting filaments. Feedback from massive stars and outflow jets disperses the cloud, limiting its star formation efficiency to a few percent.

The foregoing simulations match many observed cloud properties, but they are limited because they arise from specified initial and boundary conditions, rather than from galactic processes of cloud formation and destruction (e.g. Klessen & Glover 2016).

## 5.2. Virial parameter and star formation rate

We investigate the relation between virial parameter and star formation in galactic survey clouds with a compilation of MW star clusters associated with MCs (Lee et al. 2016; L16). This compilation estimates $\Sigma_{\text{SFR}}(R_{\text{gal}})$, the typical star formation rate surface density associated with the positions of MD17 clouds in $\Delta R_{\text{gal}} = 1$ kpc bins. For the range 4 kpc $\leq R_{\text{gal}} \leq$ 14 kpc, $\Sigma_{\text{SFR}}$ is well fit by the exponential relation

$$\Sigma_{\text{SFR}} = \Sigma_{\text{SFR},0} \exp(-R_{\text{gal}}/R_{\text{SFR},0}) \qquad (11)$$

where $\Sigma_{\text{SFR},0} = 143 \, M_\odot \, \text{pc}^{-2} \text{Gyr}^{-1}$ and $R_{\text{SFR},0} = 1.7 \pm 0.3$ kpc (L16 figure 9). A more detailed estimate of $\Sigma_{\text{SFR}}$ based on 70 μm emission also finds an approximately exponential decline, with a scale length 2.3 kpc over the same range of $R_{\text{gal}}$ (Elia et al. 2025 Figure 9).

To obtain $\alpha_{vir}$ as a function of $\Sigma_{SFR}$ we equate $R_{gal}$ in equation (11) to $R_{gal}$ in the fit equation in Figure 2, and we equate $R_{gal}$ in equation (11) to $R_{gal}$ in the fit equation in Figure 3. Then for each CO survey, $\alpha_{vir}$ and its best-fit model are each inversely correlated with $\Sigma_{SFR}$.

Figure 5 shows the anticorrelation trend between $\Sigma_{SFR}$ and each of the two CO survey estimates of $\alpha_{vir}(R_{gal})$. This trend was first noted by L16 for MD17 data. The HD15 data added here establishes this trend more clearly. This trend supports the idea that stars should form in gravitationally bound regions (L16).

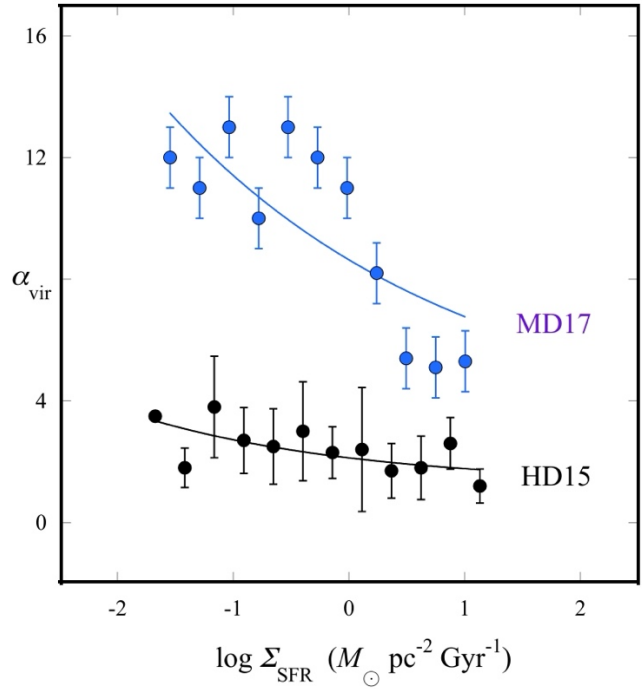

**Figure 5.** Virial parameter $\alpha_{vir}(R_{gal})$ as a function of the log of the star formation rate surface density, $\log \Sigma_{SFR}(R_{gal})$ at the same $R_{gal}$. *Circles* show values of $\alpha_{vir}$ (eq. 1) from MD17 observations in Figure 2 *(blue)* and of $\alpha_{vir}$ from HD15 observations in Figure 3 *(black)*. *Curves* show the best-fit DE model of $\alpha_{vir}$ (eq. 5) to data (eq. 1) from MD17 observations *(blue)* and from HD15 observations *(black)*. At each value of $R_{gal}$, $\Sigma_{SFR}(R_{gal})$ is obtained from eq. (11) as given by L16. Error bars on $\alpha_{vir}$ have the same values as in Figure 2 and as in Figure 3.

Figure 5 also shows that clouds whose virial parameter exceeds the critical value $\alpha_{vir} \approx 2$ have non-negligible values of star formation rate surface density. As noted above, SVE clouds with $\alpha_{\rm vir} > 2$ are expected to expand and disperse rather than to condense and form stars. In contrast the equilibrium of PVE clouds with $\alpha_{\rm vir} > 2$ is stable, so they should neither collapse nor disperse. In both SVE and PVE, such clouds with $\alpha_{\rm vir} > 2$ are not expected to globally collapse.

To estimate the number of stars that actually form in clouds with $\alpha_{\rm vir} > 2$, we find the galactocentric radius where $\alpha_{\rm vir} = 2$. According to equation (8) this radius is $R_{\rm gal, \alpha=2} = 7.3$ kpc. Thus MCs extending from the Solar neighborhood to the outer galaxy should be in stable. equilibrium if they are in PVE.

The SFR of these $\alpha_{\rm vir} > 2$ clouds is obtained by integrating equation (11) for $\Sigma_{\rm SFR}$ over $R_{\rm gal} = 7.3 - 15$ kpc, yielding $SFR_{\alpha>2} = 0.19\ M_\odot\ {\rm yr}^{-1}$. If a dense core collapses to form a star whose mass is typical of the IMF, its SFR is $SFR_{\rm core} = 2.1 \times 10^{-6}\ M_\odot\ {\rm yr}^{-1}$ as estimated in Section 5.3 below. The number of such stars born in all $\alpha_{\rm vir} > 2$ clouds in the MW is then $N_{\rm stars,\alpha>2} = (SFR_{\alpha>2}/SFR_{\rm core})(\tau_{\rm sf}/\tau_{\rm ff,core}) = 1.6 \times 10^6$ stars born during a typical star-forming cloud lifetime of $\tau_{\rm sf} = 3$ Myr. Thus more than one million stars may be expected to form in MW clouds with $\alpha_{\rm vir} > 2$, despite the apparent stability of such clouds against global collapse if they are in PVE.

It then remains to understand how such stable PVE clouds could provide significant rates of star formation in the MW. The next section describes approaches to this problem.

### 5.3. Matching the Milky Way star formation rate

Understanding the MW star formation rate ($SFR_{\rm MW}$) is a basic goal of star formation research. Many studies based on properties of young stars have found values close to $SFR_{\rm MW} = 1.9 \pm 0.4\ M_\odot\ {\rm yr}^{-1}$ (Chomiuk & Povich 2011); see also $SFR_{\rm MW} = 1.7 \pm 0.2\ M_\odot\ {\rm yr}^{-1}$ (Licquia & Newman 2015). However, estimating $SFR_{\rm MW}$ from the masses and freefall times of CO survey clouds (Goldreich & Kwan 1974) yields estimates too large by two orders of magnitude (Zuckerman & Palmer 1974, Zuckerman & Evans 1974). After unbound survey clouds are excluded, the estimated $SFR_{\rm MW}$ is still too large by a factor of ~20 (E21). A possible resolution to this problem is based on the metallicity dependence of the conversion factor $X_{\rm CO}$, and on a decrease in the star formation efficiency per free fall time $\epsilon_{\rm ff}$ with virial parameter $\alpha_{\rm vir}$. This decrease is based on fits to simulations of supersonic MHD turbulent clouds (Padoan et al. 2012; E22; Elia et al. 2025).

This section describes an alternate model, which matches the $SFR_{\rm MW}$ without assuming that MW clouds are dynamically collapsing on a free fall time scale. This model draws on the star-forming properties of dense cores in filamentary networks, based on observations of molecular clouds forming low-mass stars (Benson & Myers 1989, Lada et al. 2010 (L10), André et al. 2014 (A14)). It is also based on the STARFORGE simulations (G21, G22) described in Section 5.1.3. This model is called "rate-matching" (RM) because it specifies the MW cloud masses and star-forming ages necessary to match the $SFR_{\rm MW}$.

The term RM should not be confused with the rotation measure $R_M$, which specifies the change in polarization angle of interstellar medium gas due to the magnetic field along the line of sight (Faraday rotation).

In the RM model, dense cores are a simplified form of the gas denser than the "transition" or "threshold" column density considered most important for star formation (L10, Lada et al. 2012, Evans et al. 2014; E14). The association of dense cores, hubs and filaments is described by Myers (2009), A14, Hacar et al. (2023), and Mattern et al. (2024; M24). This model is applied here to all the star-forming clouds in the MW, not only those with $\alpha_{\rm vir} > 2$.

We use the following terminology. A protostellar dense core is dynamically unstable, collapsing, and is forming a Class 0 protostar. It is more evolved than a "starless" or "prestellar" dense core. The Class 0 protostar is in its main accretion phase (A14). Such protostars are a subset of young stellar objects (YSOs), which include the more evolved Class I protostars, flat-spectrum sources, and class II/III pre-main-sequence stars with disks (Evans et al. 2009, Großschedel et al. 2019).

We adopt dense core properties derived from observations of the Aquila molecular cloud, which was the subject of a detailed study in the *Herschel* Gould Belt survey (Könyves et al. 2015; K15). The cloud is a region of active star formation with highly filamentary structure, whose core mass distribution resembles that of the stellar initial mass function (IMF). The median of its core mean densities is $n_{\rm core} = 4 \times 10^4$ cm$^{-3}$, its core-to-star mass efficiency is $\epsilon_{\rm cs} = 0.4$, and 58 of its cores are identified as protostellar. Its prestellar and protostellar cores are selected to have $\alpha_{\rm vir} \lesssim 2$. This criterion is also satisfied in studies of dense cores in numerous star-forming regions, based on NH$_3$ line observations (e.g. Foster et al. 2009).

In infrared dark clouds and protostellar clusters forming massive OB stars, cores tend to be warmer and more massive than in Aquila. Their mass estimates are more uncertain and in some regions their mass distributions are top-heavy compared to the IMF (Pouteau et al. 2023, Motte et al. 2025). We adopt Aquila core properties here, because their masses are better determined and more representative of the IMF than in regions forming more massive stars.

In the RM model, each core forms a star whose mass equals the mean mass of the single-star IMF, $m_{star} = 0.36\ M_\odot$ (Chabrier et al. 2002, Weidner & Kroupa 2006). This choice is supported by the recent estimate $m_{star} = 0.41\ M_\odot$ from a census of over 3000 stars and brown dwarfs (Kirkpatrick et al. 2024). The initial protostellar core mass is then $m_{core} = m_{star}/\epsilon_{cs} = 0.90\ M_\odot$, its freefall time is $\tau_{ff,core} = 0.17$ Myr and its star formation rate is $SFR_{core} \equiv m_{star}/\tau_{ff,core} = 2.1 \times 10^{-6}\ M_\odot\ yr^{-1}$.

The duration of the Class 0 evolutionary phase is assumed equal to the protostellar core free fall time $\tau_{ff,core} = 0.17$ Myr. This time is consistent with the Class 0 duration estimate of 0.15 Myr from population statistics (Dunham et al. 2014), and with the time $0.12^{+0.12}_{-0.07}$ Myr during which a protostellar core is detectable in a STARFORGE simulation (O25 Table 3).

During the star-forming life of a molecular cloud, its mass $M_{MC}$ is assumed constant in time. This assumption is supported by a STARFORGE simulation of a $2 \times 10^5\ M_\odot$ cloud. Over its 3 Myr star-forming life, the cloud retains ~80% of its initial mass against dispersal by massive star feedback (G19).

The star formation rate of the MW, $SFR_{MW} = 1.9\ M_\odot\ yr^{-1}$ (Chomiuk & Povich 2011), and the MC mass of the MW, $M_{MCs,MW} = 1.0 \times 10^9\ M_\odot$ (HD15, MD17) are also assumed constant in time. To match the $SFR_{MW}$ with adopted core properties requires $N_{cores,MW} = SFR_{MW}/SFR_{core} = 9.0 \times 10^5$ protostellar cores to be distributed among the MCs of the MW. The total protostellar core mass in the MW is then $M_{cores,MW} = N_{cores,MW} m_{core} = 8.1 \times 10^5\ M_\odot$, whence the total Class 0 protostar mass in the MW is $M_{Class0,MW} = \epsilon_{cs} M_{cores,MW} = 3.2 \times 10^5\ M_\odot$. This mass is a fraction of the total mass of MCs in the MW, equal to the MW "Class 0 efficiency" $\mu_{Class0,MW} \equiv M_{Class0,MW}/M_{MCs,MW} = 3.2 \times 10^{-4}$.

### 5.3.1. Protostars in a molecular cloud

The number of Class 0 protostars in a MC of mass $M_{MC}$ is assumed to equal the mass-weighted fraction of the number of Class 0 protostars needed to match the $SFR_{MW}$. Thus

$\mu_{\text{Class 0,MC}} = \mu_{\text{Class 0,MW}}$ or $N_{\text{Class 0,MC}} = 0.0093(M_{\text{MC}}/M_\odot)$. Here $M_{\text{MC}}$ is the cloud mass above a minimum column density consistent with MW CO surveys.

This number $N_{\text{Class0,MC}}$ is assumed to be in steady state during the cloud star-forming age, denoted $\tau_{\text{sf}}$. In this steady state the birth rate of Class 0 protostars $N_{\text{Class0,MC}}/\tau_{\text{ff,core}}$ equals the rate of their transition from evolutionary Class 0 to Class I. With these assumptions, the number of Class 0 protostars in a MC is constant in time and is linearly proportional to $M_{\text{MC}}$, according to

$$N^{\text{RM}}_{\text{Class0,MC}} = \frac{SFR_{\text{MW}}\, M_{\text{MC}}}{SFR_{\text{core}}\, M_{\text{MCs,MW}}} \quad . \tag{12}$$

The corresponding protostellar mass fraction of the MC, or its Class 0 protostar efficiency, is $SFE_{\text{Class 0}} = \tau_{\text{ff,core}}(SFR_{\text{MW}} / M_{\text{MCs,MW}}) = 3.2 \times 10^{-4}$. It is independent of cloud mass, as discussed in Section 6.2.5.

**5.3.2. Young stellar objects in a molecular cloud** The cumulative number of YSOs in a MC equals the number of Class 0 protostars born in each "generation" of duration $\tau_{\text{ff,core}}$, multiplied by the number of generations in the cloud's star-forming age $\tau_{\text{sf}}/\tau_{\text{ff,core}}$. Then the RM number of YSOs is given by equation (12) multiplied by $\tau_{\text{sf}}/\tau_{\text{ff,core}}$, or

$$N^{\text{RM}}_{\text{YSOs,MC}} = \frac{SFR_{\text{MW}}\, M_{\text{MC}}\, \tau_{\text{sf}}}{SFR_{\text{core}}\, M_{\text{MCs,MW}}\, \tau_{\text{ff,core}}} \quad . \tag{13}$$

In equation (13), $N^{\text{RM}}_{\text{YSOs,MC}}$ is proportional to the product of the cloud mass and the cloud age, assuming that all YSOs born in the cloud remain within its observational boundary. Equation (13) can be used to draw isochrones in a plot of observed values of $\log N_{\text{YSOs}}$ vs. $\log M_{\text{MC}}$ as in Figure 6.

**5.3.3. Molecular cloud star and core formation rates**

The formation rate of Class 0 protostars in a MC is equal to the formation rate of YSOs in the same MC, as equations (12) and (13) show. Thus $SFR^{\text{RM}}_{\text{Class0,MC}} = m_{\text{star}} N^{\text{RM}}_{\text{Class0,MC}}/\tau_{\text{ff,core}}$

and $SFR^{RM}_{YSOs,MC} = m_{star} N^{RM}_{YSOs,MC}/\tau_{sf}$. Each of these rates is equal to the *SFR* of the MC,

$$SFR^{RM}_{MC} = \frac{SFR_{MW}}{M_{MCs,MW}} M_{MC} \quad . \tag{14}$$

The corresponding formation rate of protostellar core mass is $CFR^{RM}_{MC} = SFR^{RM}_{MC}/\epsilon_{cs}$, or

$$CFR^{RM}_{MC} = \frac{SFR_{MW}}{\epsilon_{cs} M_{MCs,MW}} M_{MC} \quad . \tag{15}$$

Equation (14) predicts that a cloud of mass $M_{MC} = 2 \times 10^4 \, M_\odot$ has a star formation rate $SFR^{RM}_{MC} = 4 \times 10^{-5} \, M_\odot \, yr^{-1}$. This rate lies within a factor ~2 of the mean rate in a STARFORGE cloud of the same mass, $SFR^{O25}_{MC} = 8 \times 10^{-5} \, M_\odot \, yr^{-1}$ (O25 Table 3), assuming that the stellar mass is 0.36 $M_\odot$ as adopted in Section 5.3.

### 5.3.4. Star formation efficiency

The observed "star formation efficiency" of a MC is denoted as

$$SFE^{obs}_{MC} = \frac{m_{star} N^{obs}_{YSOs,MC}}{M_{MC}} \tag{16}$$

as given in E21. The predicted RM value $SFE^{RM}_{sf,MC}$ is derived from the RM number of YSOs in equation (13), as

$$SFE^{RM}_{sf,MC} = \frac{SFR_{MW} \tau_{sf}}{M_{MCs,MW}} \quad . \tag{17}$$

Equation (17) predicts that the star-forming efficiency of a MC is independent of its mass and galactic radius, but increases linearly with its star-forming age, $SFE^{RM}_{sf,MC} = 0.0019(\tau_{sf}/Myr)$. If a cloud has star-forming age $\tau_{sf} = 3 - 10$ Myr, it has $SFE^{RM}_{sf,MC} = 0.006 - 0.02$. This range is similar to that observed in local clouds (Pokhrel et al. 2020; P20). Similarly, if the MC has density 100 cm$^{-3}$ its free fall time is $\tau_{ff}$ = 3.3 Myr, and its star formation efficiency per free-fall time (Krumholz et al. 2014) is $SFE_{ff,MC}^{RM} = 0.0019(\tau_{ff}/\text{Myr}) = 0.006$ as given in E22.

The linear increase of $SFE$ with star-forming age in equation (17) is also approximated in some (but not all) cloud evolution simulations. In G19 Figure 2 the cloud $SFE$ increases nearly linearly from an initial value of ~0.01 % up to ~1 % after 3 Myr. This linear behavior is seen after an initial superlinear increase and before a sublinear stage where the $SFE$ levels off due to cloud dispersal. More nearly-linear $SFE$ histories are seen in a study where self-gravity is opposed progressively by the combination of turbulence, magnetic fields, outflow jets, and radiative heating (Appel et al. 2023; A23). In general, the dependence of $SFE$ on age in simulations depends on the definition of the star-forming age of the cloud and on the adopted forms of feedback.

### 5.3.5. Comparison with observations

The foregoing RM model predictions are compared with observations of local clouds in two ways. The model predicts the number of Class 0 protostars, the number of YSOs, and the star formation efficiency, provided a cloud star-forming age is well determined from YSO pre-main sequence (PMS) evolutionary tracks. This comparison is made for the Orion A cloud. For a group of clouds with less certain PMS ages, the model predicts the distribution of their star-forming ages. This second comparison applies to estimates of the mass and number of YSOs in eleven star-forming clouds within 500 pc of the Sun (Lada et al. 2010; L10).

For each comparison, it is necessary to choose a minimum cloud surface density used to estimate $N_{YSO}$ and $M_{MC}$. This surface density should be consistent with observations of MW CO clouds, since equation (13) depends on the total molecular mass derived from CO observations. An observed minimum CO cloud surface density is estimated from a summary of MW CO observations, where $\Sigma_{cld,min} \approx 12\ M_\odot\ \text{pc}^{-2}$ (HD15 Figure 8). The corresponding near-infrared extinction $A_K \approx 0.08$ mag is nearly equal to the boundary level $A_K = 0.1$ mag used for cloud mass and YSO estimates by L10. Therefore we use the L10 estimates of $N_{YSO}$ and $M_{MC}$ with $A_K = 0.1$ mag for comparisons with the RM model.

A detailed study of YSOs in the Orion region was made using the DBSCAN algorithm to identify separate clusters, the PARSEC library of stellar evolutionary tracks to estimate PMS ages, and radial velocities, distances, and proper motions to estimate expansion ages (Zari et al. 2019; Z19). Of the three groups projected on the Orion A cloud, groups C and D are considered to be associated with the Orion A cloud. Group C has 943 members with PMS age $7.9^{+0.6}_{-0.2}$ Myr and expansion age $8^{+0.5}_{-0.04}$ Myr, while group D has 60 members, PMS age $7.1^{+0.5}_{-0.3}$ Myr, and expansion age $7^{+0.6}_{-0.2}$ Myr (Z19 Table 2). Based on these findings, we take the star-forming age of the Orion A cloud to be the age of its oldest stars, $8.0 \pm 0.5$ Myr.

With an age estimate of 8 Myr from Z19 and cloud mass from L10, the observed and predicted young star populations and efficiencies in the Orion A cloud agree closely. The observed numbers of Class 0 protostars and YSOs are $N^{obs}_{Class0} = 60$ and $N^{obs}_{YSO} = 2980$ (Großschedel et al. 2019). The predicted numbers are $N^{RM}_{Class0} = 64$ and $N^{RM}_{YSO} = 2860$ from equations (12) and (13). The observed and predicted efficiencies are $SFE^{obs}_{MC} = 0.015$ and $SFE^{RM}_{MC} = 0.015$ from equations (16) and (17). It will be important to make similar comparisons as more detailed studies of massive star-forming clouds become available.

The RM number of YSOs in a cloud of mass $M_{MC}$ at star-forming age $\tau_{sf}$ are shown as four isochrone lines in Figure 6, based on equation (13). Each isochrone line predicts $\log N_{YSO}$ as a function of $\log M_{MC}$ at an age $\tau_{sf} = 1, 2, 4,$ or 8 Myr. These lines are superposed on data points representing the observed number of YSOs in local molecular clouds of observed mass $M_{MC}$. The points are based on YSO counts of *Spitzer Space Telescope* observations and on near-infrared extinction maps, with boundary value $A_K = 0.1$ mag (L10).

The placement of the RM isochrone pattern in Figure 6 has estimation uncertainty due to uncertainties in the MW parameters $M_{MCs,MW}$ and $SFR_{MW}$ in equation (13). The relative uncertainties of 0.30 in $M_{MCs,MW}$ and 0.21 in $SFR_{MW}$ imply that the RM prediction of $N_{YSO}$ for a cloud of fixed mass and age has relative uncertainty $\sigma_{N_{YSO}}/N_{YSO} = 0.37$. In the logarithmic plot of Figure 6 the corresponding sigma is $\sigma_{\log N_{YSO}} = (1/\ln 10)(\sigma_{N_{YSO}}/N_{YSO}) = 0.16$. Thus the pattern of the four $\log N_{YSO}$ isochrones may shift vertically with respect to the cloud data points within a one-sigma range of $\pm 0.16$ due to uncertainty in $M_{MCs,MW}$ and $SFR_{MW}$.

The distribution of cloud data points in Figure 6 is fairly consistent with the predictions of the RM model. The average predicted YSO age over the 11-cloud sample is $\overline{\tau_{sf}}/2 \approx 2$ Myr. This age approximates the duration of Stage II, the most common YSO evolutionary stage (Evans et al. 2009). Half of the clouds in Figure 6 have age predictions within ~1 Myr of their stellar age estimates, based on available pre-main sequence (PMS) model isochrones. These age estimates span ~1 Myr to ~10 Myr. Assuming that the uncertainty in $\tau_{sf}^{RM}$ is ~ 1 Myr, the closest matches are in Orion A, with $\tau_{sf}^{RM} = 8$ Myr and $\tau_{sf}^{PMS} = 8 \pm 0.5$ Myr (Z19), in Orion B, with ~2 Myr and ~2 Myr (Flaherty & Muzerolle 2008), and in the California cloud, with ~1 Myr and ~1 Myr (Wolk et al. 2010). Considering the uncertainties in $\tau_{sf}^{RM}$, the age estimates for the Pipe and California clouds are each rounded from 0.5 Myr to 1 Myr.

Despite the similar locations of the Orion A and Orion B regions, the available PMS and RM age estimates each indicate that the oldest stars in Orion A are significantly older than the oldest stars in Orion B.

The vertical spread of log $N_{YSO}^{obs}$ data points in Figure 6 is consistent with the RM model only if the cloud star-forming ages $\tau_{sf}$ also have a significant range of values. If all of the clouds had the same age, the range of values of $N_{YSO}^{obs}$ would significantly exceed the range expected from estimation error alone. Instead, consistency with $N_{YSO}^{obs}$ values requires that $\tau_{sf}$ have a range of ages for each of three groups of clouds having nearly the same mass within each group. These ranges have mean ± standard deviation $\tau_{sf} = 5 \pm 4$ Myr for the low-mass clouds RCrA, Lup 3, Lup 1, and Lup 4. They have $\tau_{sf} = 4 \pm 3$ Myr for the medium-mass clouds Pipe, Oph, Tau, and Per, and $\tau_{sf} = 3 \pm 3$ Myr for the high-mass clouds Cal, Ori B, and Ori A. These ranges can be verified by inspection of Figure 6.

The clouds in Figure 6 have star formation efficiency at the current star-forming age $SFE_{sf} = 0.001 - 0.015$ with median 0.008 (L10). These properties are similar to those in an independent sample of 12 local clouds, having $SFE_{sf} = 0.005 - 0.016$, with median 0.010 (Pokhrel et al. 2020; P20). This P20 sample was also analyzed to estimate the star formation efficiency in a free fall time, yielding $SFE_{ff} = 0.01 - 0.04$, with median 0.026 (Pokhrel et al. 2021).

When the L10 and P20 samples are combined without duplication, their 20 clouds have a distribution of ages $\tau_{sf}$ with quartile and median values 2.9, 4.2, and 8.0 Myr. These ages are exemplified by the Lupus 1, Perseus, and Orion A clouds, each with age uncertainty ~1 Myr.

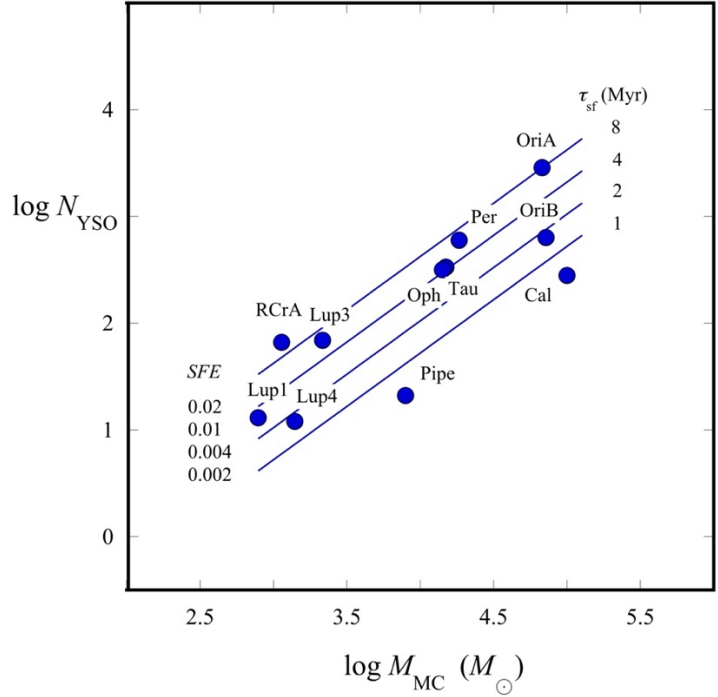

**Figure 6.** Comparison of observed and predicted $N_{YSO}$, the number of YSOs in 11 molecular clouds within 500 pc of the Sun, which have mass $M_{MC}$ estimated from observations. *Filled circles* show observed values of $N_{YSO}$ based on near-infrared extinction maps with $A_K \geq 0.1$ mag (L10 Table 2). *Lines* show RM model isochrones for cloud star-forming ages $\tau_{sf} = 1 - 8$ Myr based on equation (13), and for values of the observed star formation efficiency $SFE = 0.002 - 0.02$. The one-sigma uncertainty in each isochrone due to uncertainties in $M_{MCs,MW}$ and $SFR_{MW}$ is 0.16.

The free fall times for the combined L10 and P20 samples are consistently greater than their star-forming ages, by a factor $\gtrsim 2$. The ratio $\tau_{ff}/\tau_{sf}$ has quartile and median values 1.6, 2.4, and 3.6-9.0. While these free fall times and star-forming ages have the same order of magnitude 2-9 Myr, their significant differences show that simple free fall collapse does not limit the evolution of local clouds.

The Class 0 protostar mass fraction predicted by equation (12) has been compared with observations in similar fashion to the YSO comparison in Figure 6. The Class 0 protostars in Perseus clumps were compared with the masses of their host clumps (Sadavoy et al. 2014, Pezzuto et al. 2021). As in Figure 6 the number of Class 0 protostars in a clump is correlated with the clump mass, with correlation coefficient 0.7 - 0.9 depending on the clump boundary definition. This comparison in Perseus is more limited by statistics than in the Orion A cloud, because most Perseus clumps have very few Class 0 protostars while Orion A has ~60 Class 0 protostars.

The foregoing comparisons with Solar neighborhood cloud data are consistent with the RM assumption that the $SFR$ of a MC is equal to the $SFR$ of the MW, weighted by the mass fraction of the MC to the molecular mass of the MW. Here we show that this assumption is also consistent with observational estimates of MC mass and of MC $SFR$ in radial rings throughout the MW.

The ring-average mass of a MC increases by a factor ~10 from outer-galaxy rings at 14 kpc to inner-galaxy rings at 4 kpc (MD17 Figure 9). At each $R_{gal}$ the ring-average surface densities $\Sigma_{SFR,MC}$ and $\Sigma_{MC}$ have nearly the same ratio from ring to ring, since their profiles $\Sigma_{SFR,MC}(R_{gal})$ and $\Sigma_{MC}(R_{gal})$ decline with nearly the same shape and scale length (L16, E25). Thus from an outer ring to an inner ring, the average $SFR_{MC}$ and the average $M_{MC}$ each increase by nearly the same ratio, provided that the MCs which have typical $SFR_{MC}$ and typical $M_{MC}$ have similar ring area filling factors.

These properties imply that the ratio of $SFR_{MC}$ to $M_{MC}$ from ring to ring is quantitatively consistent with the ratio of total MW $SFR$ to total MW $M_{MC}$. For $R_{gal} = 4$ to 14 kpc, the observed surface densities can be expressed as $\Sigma_{SFR,MC} = \Sigma_{SFR,0}\exp(-R_{gal}/R_{SFR,0})$ and $\Sigma_{MC} = \Sigma_{MC,0}\exp(-R_{gal}/R_{MC,0})$ (L16 Figure 9). When $R_{SFR,0} \approx R_{MC,0}$ the ratio $SFR_{MC}/M_{MC}$ can be approximated as $SFR_{MC}/M_{MC} \approx \Sigma_{SFR,MC}/\Sigma_{MC} \approx \Sigma_{SFR,0}/\Sigma_{MC,0}$, independent of $R_{gal}$. This property allows an estimate of $SFR_{MW}$ using equation (14) and the values $\Sigma_{SFR,0} = 143\ M_\odot pc^{-2} Gyr^{-1}$ and $\Sigma_{MC,0} = 95\ M_\odot pc^{-2}$ (L16). Then $SFR_{MW} = (\Sigma_{SFR,0}/\Sigma_{MC,0})M_{MCs,MW}$ where $M_{MCs,MW}$ is the total molecular cloud mass of the MW, $M_{MCs,MW} = (1.0 \pm 0.3) \times 10^9\ M_\odot$ (HD15) as in Section 5.3. The resulting star formation rate is $SFR_{MW} = 1.5 \pm 0.5\ M_\odot\ yr^{-1}$. This estimate is based on MW survey data for $R_{gal} > 4$ kpc. Nonetheless it is consistent with earlier estimates based on a wider range of $R_{gal}$, $SFR_{MW} = 1.9 \pm 0.4\ M_\odot\ yr^{-1}$ (Chomiuk & Povich 2011), $1.7 \pm 0.2\ M_\odot\ yr^{-1}$ (Licquia & Newman 2015) and $1.4^{+0.6}_{-0.4}$ based on 70 μm emission (E25).

In summary, when YSO ages are known with relatively good precision as in Orion A, the RM model predicts with good accuracy the observed numbers of Class 0 protostars and YSOs, and the star formation efficiency. For local clouds with less certain ages, the RM model requires a range of cloud ages over 1-10 Myr and mean YSO age ≈ 2 Myr for consistency with observed cloud masses and numbers of YSOs. For Class 0 protostars in Perseus, the number of protostars in each clump is correlated with their surrounding clump mass, as expected from the RM assumption $\mu_{\text{Class 0,MC}} = \mu_{\text{Class 0,MW}}$. Observed MC masses and star formation rates in radial galactic rings are consistent with this assumption of equal mass fractions, and with standard estimates of the MW $SFR$.

The increase in $N_{\text{YSO}}$ with star-forming cloud age implied by the RM model may also be consistent with the increase in $N_{\text{YSO}}$ with cloud dense gas fraction identified in local clouds (L10, E14). The star-forming age and the dense gas fraction may be correlated if a typical MC develops more accreting core-forming filaments as it ages. The fragmentation of filaments into cores is discussed in H23 and Chira et al. (2018). This possible combination of properties is a good subject for further investigations, with observations and simulations.

# 6. Discussion

## 6.1. Limitations

This section describes limitations of the foregoing analysis and models in Sections 2-5.

### 6.1.1. Elongated and magnetically subcritical clouds

The typical MC properties in this paper do not describe the properties of all MCs in the MW. MW "bones" (Zucker et al. 2015) are filamentary star-forming molecular clouds in the inner MW. A catalog of 45 large-scale filaments indicates typical mass $\sim 10^4\ M_\odot$, length ~30 pc, and aspect ratio ~20 (Zucker et al. 2018). These clouds resemble local clouds in Figure 6 in their masses and in their level of star formation, since their median $SFE$ is ~ 0.01 after correction for differences in distance (Zhang et al. 2019). The majority of star formation in the MW may occur in filamentary structures (Pillsworth & Pudritz 2024). *SOFIA* polarization observations of ten highly elongated filaments indicate that their magnetic fields have complex, partially ordered structure. Their typical mass to magnetic flux ratio is magnetically subcritical ($\lambda < 1$) in contrast to most molecular cloud estimates, approaching $\lambda \approx 1$ in their star-forming regions (Stephens et al. 2022, Coudé et al. 2025, Stephens et al. 2025).

### 6.1.2. DE and PVE models

Molecular clouds in the MW are described in Section 2 as belonging to a pressure-bounded virial equilibrium system in "dynamical equilibrium" (DE) discussed by OK22. The PVE model used here relies only on the pressure balance between midplane clouds and the gravitational weight of surrounding gas and nearby stars expressed in OK22 equation (7). It does not rely on the balance between heating and cooling rates and other features of the OK 22 DE simulations, including feedback from massive stars and their supernovae. Thus the successful fitting of OK22 equation (7) to CO survey data should be considered a verification of only one aspect of the DE model.

### 6.1.3. Virial parameter values

The critical values of the virial parameter in the SVE and PVE models in Section 5.1 are uncertain because possible variations in cloud elongation, central concentration, and magnetization are not known for each cloud. As noted earlier the expected departure of $\alpha_{\text{vir}}$ from its value for the uniform spherical case is by a factor $\lesssim 2$. Thus a range of critical values near $\alpha_{\text{vir}} \approx 2$ divides unbound from bound SVE clouds, or unstable from stable PVE clouds. Nonetheless it remains clear that many survey clouds exceed values in the critical range. Thus many CO survey clouds are either unbound according to SVE, or their equilibrium is stable according to PVE, or their dynamical status is neither equilibrium nor simple global collapse.

### 6.1.4. Protostellar core properties

Protostellar core masses are assumed to have a single constant value, producing stars having a single value of mass, equal to the mean mass of the IMF. This simplification ignores the major role of massive stars and their feedback in cloud evolution and dispersal. It is nonetheless justified for the purpose of matching the formation rate of MW stars. Due to the strongly peaked nature of the IMF, the mean, median and mode of the IMF all have values less than 1 $M_\odot$ (Maschberger 2013). Thus the galactic star formation rate can be reasonably approximated by the formation rate of its low-mass stars.

### 6.1.5. CO conversion factor

The metallicity of MW gas, normalized to Solar neighborhood values, is expected to decrease from $Z \approx 1.5$ to $0.5$ as $R_{gal}$ increases from 4 to 14 kpc (E22, Figure 1). Such a decrease in Z has been proposed to cause an increase in the conversion factor $X_{CO}$ from $X_{CO} = 2.1$ to $3.1$ (Lada & Dame 2020), or from $X_{CO} = 1.5$ to $3.0$ (E22) over the same range of $R_{gal}$. These factors of increase in $X_{CO}$ are $F_X = X(14)/X(4) = 1.5 - 2.0$. They are significantly smaller than the factors of decrease in surface density $F_\Sigma$ based on the two CO surveys in Figure 1. These factors are derived by assuming the standard value of $X_{CO} = 2.0 \times 10^{20}$ cm$^{-2}$ (K km s$^{-1}$)$^{-1}$(Bolatto et al. 2013). They can be written $F_\Sigma = \Sigma(4)/\Sigma(14) = \exp[10/(R_0/\text{kpc})]$ based on the exponential fits in Figure 1. Here $R_0$ is the scale length of each exponential fit. Then $F_\Sigma \approx 10$ for the HD15 data and $F_\Sigma \approx 17$ for the MD17 data.

In summary, the variation in $X_{CO}$ over $R_{gal}$ = 4 to 14 kpc due to metallicity appears significant (factors of 1.5-2), but insufficient to account for the observed declines in MC surface density (factors of 10-17). Consequently we neglect the variation of metallicity in evaluating the variation of cloud surface density over $R_{gal}$.

### 6.1.6. Protostar formation rate

A key feature of the RM model is the requirement that a MC of mass $M_{MC}$ form protostars at a rate $SFR_{MC}^{RM} \approx 10^{-3}(M_{MC}/M_\odot)$ $M_\odot$ Myr$^{-1}$ due to collapsing protostellar cores during its star-forming lifetime. This rate agrees within a factor ~2 of the mean rate in the STARFORGE simulation of O25, as noted in Section 5.3.3. However our understanding of how protostellar cores develop remains incomplete. More detailed simulations are needed to study the growth rates of core-forming filaments, as in Chira et al. (2019). More observational studies are also needed of the gas motions associated with cores and their filaments, as in the Perseus molecular clouds, as in Chen et al. (2024).

### 6.1.7. MC Core mass fractions

The RM model appears consistent with observations of YSOs in local clouds as discussed in Section 5.3.5. However it remains to compare model predictions with the incidence of YSOs in more distant clouds, and it remains to compare model predictions with the incidence of Class 0 protostars in both local and distant clouds. Protostars are far fewer in number than YSOs, so insufficient sensitivity and incompleteness may limit estimation accuracy. Nonetheless it should be possible to estimate protostar mass fractions, or upper limits on their mass fractions, in well-studied regions.

## 6.2. Implications

The following points are the main implications of this work:

### 6.2.1 Virial parameter increases with galactocentric radius

Each of two CO surveys reveals a trend of virial parameter $\alpha_{\text{vir}}$ increasing by a factor ~2 from $R_{\text{gal}} = 4$ to 14 kpc. This trend is inconsistent with MW clouds in SVE. It is consistent with MW clouds in PVE, if (*a*) the midplane pressure on MCs is due mainly to the gravitational weight of nearby stars, and (*b*) the surface densities of stars and of molecular clouds have approximately exponential declines with $R_{\text{gal}}$. The best-fitting trend of $\alpha_{\text{vir}}(R_{\text{gal}})$ is summarized in equation (8). It indicates that inner galaxy clouds are more strongly self-gravitating and have greater pressure ratio $P_{\text{int}}/P_{\text{ext}}$ than outer galaxy clouds, each by a factor ~2.

### 6.2.2. Environmental pressure on Milky Way clouds is typical of disk galaxies

The median virial parameter of MW clouds is similar to that in S22, indicating that MW clouds have virial parameter values typical of a large sample of disk galaxies. This result extends the evidence of environmental influence on MCs in disk galaxies (Leroy et al. 2008, Barrera-Ballesteros et al. 2021, Sun et al. 2022 (S22), L24, L25, Schinnerer & Leroy 2024).

### 6.2.3 PVE clouds with $\alpha_{\text{vir}} > 2$ form stars without global collapse

Survey clouds with $\alpha_{\text{vir}} > 2$ are known to form significant numbers of stars. But their consistency with PVE implies that they cannot form stars by global collapse. This apparent contradiction may be resolved if cloud structure approximates conditions of PVE on the large scale and of collapsing protostellar cores in filaments on the small scale. These properties resemble those in simulations of nearly critical clouds whose initial turbulence dissipates on a crossing time scale (G19). These clouds undergo minimal global contraction, while forming filaments, cores and protostars.

### 6.2.4 A "bottom-up" model of star formation

In the rate-matching (RM) model, the $SFR_{MW}$ is ascribed to numerous small-scale collapses, each involving a small fraction of the cloud mass and a small fraction of the cloud lifetime. This picture follows the "threshold" or "transition" concept where only gas denser than a minimum density tends to forms stars (L10, Lada et al. 2012, E14, M24). It is motivated by studies of dense cores having masses $\approx 1\ M_\odot$, density $\approx 10^4$ cm$^{-4}$, and free fall time $\tau_{ff} \approx 0.1$ Myr. They are numerous in well-studied MCs (K15) and may exemplify initial conditions for low-mass star formation (Shu et al. 1987, H23).

In this model a MC of mass $M_{MC}$ forms new protostellar cores at the rate $CFR_{MC} = 0.0048(M_{MC}/M_\odot)\ M_\odot$ Myr$^{-1}$ in order to match the $SFR_{MW}$. New cores are expected to form as the MC filamentary network grows. The model time scales are the core free fall time and the MC star-forming age, which is likely set by outflows and massive star feedback (Chevance et al. 2020). The global free fall time of the cloud may not be relevant, since the cloud is not expected to collapse in global free fall.

### 6.2.5. Rate-matching model in relation to MW structure and cloud virial parameter

The RM assumptions that Class 0 protostars have equal mass fractions and constant formation rate are supported by observations of protostars and YSOs in Solar neighborhood clouds, and by estimates of ring-average MC mass and $SFR$ over $R_{gal} = 4 - 14$ kpc, described in Section 5.3.5. The similar dependence of $\Sigma_{SFR}$ on $R_{gal}$ and of $\Sigma_{MC}$ on $R_{gal}$ (L16, E25) also implies that MCs have ring-average efficiencies $SFE_{sf}$ and $SFE_{ff}$ which are independent of $M_{MC}$ and of $R_{gal}$. The G19 MHD simulation study found that the true $SFE$ of GMCs has no strong trend with GMC mass.

One may question how the observed increase of $SFR$ toward the inner galaxy (L16) can be consistent with the above RM property that $SFE$ is independent of $R_{\text{gal}}$. These properties may be consistent because inner galaxy clouds are more massive than outer galaxy clouds by an order of magnitude, as noted in Section 5.3.5. This property is also evident for the typical *PHANGS* galaxy, where the molecular gas surface density on 150 pc scales increases by a factor ~30 from outer to inner radii (S20). Thus inner galaxy clouds may have a greater $SFR$ than outer galaxy clouds because they have similar $SFE$ but more mass, rather than having greater $SFE$ but similar mass.

On the other hand, the CO survey analysis of Section 4 indicate that $\alpha_{\text{vir}}$ increases with $R_{\text{gal}}$. If the $SFE$ of a cloud is independent of $R_{\text{gal}}$ as found above, these results imply that more strongly bound, massive clouds in the inner galaxy have about the same $SFE$ as less massive, weakly bound clouds in the outer galaxy, provided the outer galaxy clouds are primarily molecular and not dispersing. This conclusion is consistent with the idea that the $SFR$ of a MC is dominated by cores and filaments on scales of 0.1 - 1 pc. These may couple only weakly to the dynamical state of the MC on scales of ~10 pc. A similar point was made about the weak correlation between $\alpha_{\text{vir}}$ and $SFE$ in observational studies (G19).

Overall, the DE model gives a useful description of pressurized MCs in global equilibrium, and the RM model offers a simple way to estimate the populations of forming stars in MCs, consistent with the $SFR_{\text{MW}}$. It remains to link the global properties of MCs in the DE model to the local properties of the protostellar cores in the RM model. That will be an important challenge for further simulations and observations.

### 6.2.6 Why is star formation inefficient?

The typical mass fraction of YSOs in MCs, $SFE_{\text{MC}} \approx 0.01$, is considered evidence of MC inefficiency in forming stars. This inefficiency may arise from outflows and/or massive star feedback, which disperse cloud gas before it can form more stars (G19, Chevance et al. 2020). In the RM model an additional limit on $SFE_{\text{MC}}$ is due to the low dense gas fraction of the MC, i.e. the low mass fraction $\approx 10^{-3}$ of its protostellar cores. This low mass fraction is probably limited by the growth rate of core-forming filaments as a cloud evolves.

## 7. Summary and conclusions

This paper describes observations and models of star formation in molecular clouds (MCs) in the Milky Way (MW). The gravitational binding of MCs becomes weaker from the inner to the outer MW, since the MCs fit models of PVE where the surface density of MW stars and gas declines with galactocentric radius. Many such MCs form stars in filaments and cores without evidence of global free fall, as predicted by simulations of initially critical clouds whose initial turbulence dissipates in a crossing time. These MCs may match the star formation rate of the MW if collapsing protostellar cores comprise ~$10^{-3}$ of their mass. A model of this process predicts ages and star formation efficiencies, in good agreement with estimates in local MCs. The main points are:

1. Two CO surveys of Milky Way molecular clouds (MCs) were analyzed to estimate the influence of MW environment on MCs as found in studies of external disk galaxies. The MC surface density $\Sigma_{\text{cld}}$ and virial parameter $\alpha_{\text{vir}}$ were analyzed in rings of width 1 kpc over $R_{\text{gal}} = 4 - 14$ kpc. The virial parameter increases with $R_{\text{gal}}$ by a factor ~2 in each survey.

2. The increasing trends of $\alpha_{\text{vir}}(R_{\text{gal}})$ in the two surveys (HD15 and MD17) are nearly identical apart from a multiplying factor of ~4, with mean values of $\alpha_{\text{vir}}$ of 2.2 (HD15) and 9.6 (MD17). This factor ~4 is attributed to differences in cloud boundary definition and observational resolution. A combined trend centered on the lower values of $\alpha_{\text{vir}}$ is adopted, since then the typical value of $\alpha_{\text{vir}}$ is more consistent with typical values in the *PHANGS-ALMA* CO survey of disk galaxies.

3. The $\alpha_{\text{vir}}(R_{\text{gal}})$ trend is fit with a model of MCs in pressure-bounded virial equilibrium (PVE), based on the dynamical equilibrium equations of OK22. The MC environment of surrounding gas and nearby stars exert pressure on the MC. The PVE of the MC is maintained by the combination of self-gravity and external pressure. A model without external cloud pressure does not fit the data.

4. The PVE model fits match the increasing trend of $\alpha_{\text{vir}}(R_{\text{gal}})$ provided the surface density of MC gas declines with $R_{\text{gal}}$ in similar form to the decline of surface density of nearby stars. The main contribution to the $\alpha_{\text{vir}}(R_{\text{gal}})$ trend is due to nearby stars. If the stellar contribution is removed, the model does not fit the data.

5. The MC virial parameter is inversely correlated with the MC star formation rate surface density due to massive star clusters. This result indicates that star formation is favored in MCs with stronger gravitational binding, as suggested by L16.

6. Many MCs from $R_{\text{gal}} = 4$ to $15$ kpc have $\alpha_{\text{vir}} \gtrsim 2$. Consistency with PVE implies that these MCs cannot form stars by global cloud collapse in a free fall time. Nonetheless they are associated with abundant star formation (SF). Consistency with both PVE and SF may require these clouds to resemble PVE on the cloud scale, but to form their stars on smaller scales without global free fall.

7. Some relevant /8MHD turbulent simulations with initial $\alpha_{\text{vir}} = 2$ form filaments, cores, and protostars with minimal global collapse motions, if their initial turbulence is allowed to dissipate on a cloud crossing time scale (STARFORGE simulations, G19, O25).

8. A "rate-matching" (RM) model matches the *SFR* of the MW and typical values of *SFE* without global cloud collapse. Its clouds maintain $\sim 10^{-3}$ of their mass in the form of collapsing low-mass cores. The masses and durations of its clouds and cores are similar to those in G19 and O25.

9. In the RM model, each MC forms YSOs in proportion to its mass and to its star-forming age $\tau_{\text{sf}}$. A sample of 20 local clouds have a spread of estimated ages with quartile values 3, 4, and 8 Myr. These estimates lie within ~1 Myr of PMS estimates in Orion A, Orion B, and the California MC. The typical ratio of free fall time to star-forming age is ~2.4. The Orion A cloud is predicted to have ~60 Stage 0 protostars and ~2900 YSOs, with $SFE \approx 0.02$, each close to observed estimates. Observed MC masses and star formation rates in radial galactic rings are also shown to be consistent with the RM model and with standard estimates of the MW *SFR*.

10. The prediction that the number of YSOs in a MC increases with cloud age may be consistent with the correlation between YSOs and dense gas fraction (L10, E14). If so, the dense gas fraction in a MC may increase with age as the cloud develops more mass in its core-forming filaments.

**Acknowledgements**

We thank Neal Evans, Charlie Lada, and Ralf Klessen for their helpful comments and advice, and Riwaj Pokhrel for useful data. We thank the referee for a prompt and helpful review. PCM acknowledges Irwin Shapiro and Terry Marshall for their support.